\documentclass[twocolumn,english,floatfix,superscriptaddress,prl,showpacs]{revtex4-1}
\usepackage[T1]{fontenc}
\usepackage[utf8]{inputenc}
\usepackage{color}
\usepackage{graphicx}

\makeatletter

\@ifundefined{textcolor}{}
{%
 \definecolor{BLACK}{gray}{0}
 \definecolor{WHITE}{gray}{1}
 \definecolor{RED}{rgb}{1,0,0}
 \definecolor{GREEN}{rgb}{0,1,0}
 \definecolor{BLUE}{rgb}{0,0,1}
 \definecolor{CYAN}{cmyk}{1,0,0,0}
 \definecolor{MAGENTA}{cmyk}{0,1,0,0}
 \definecolor{YELLOW}{cmyk}{0,0,1,0}
}

\renewcommand\[{\begin{equation}}
\renewcommand\]{\end{equation}}

\makeatother

\usepackage{babel}
\begin{document}

\title{Strong-disorder renormalization group study of the Anderson localization
transition in three and higher dimensions}

\author{H. Javan Mard}

\affiliation{Department of Physics and National High Magnetic Field Laboratory,
Florida State University, Tallahassee, FL 32306}

\author{José A. Hoyos}

\affiliation{Instituto de Física de São Carlos, Universidade de São Paulo, C.P.
369, São Carlos, SP 13560-970, Brazil}

\author{E. Miranda}

\affiliation{Instituto de Física Gleb Wataghin, Unicamp, R. Sérgio Buarque de
Holanda, 777, Campinas, SP 13083-859, Brazil}

\author{V. Dobrosavljević}

\affiliation{Department of Physics and National High Magnetic Field Laboratory,
Florida State University, Tallahassee, FL 32306}
\begin{abstract}
We implement an efficient strong-disorder renormalization-group (SDRG)
procedure to study disordered tight-binding models in any dimension
and on the Erdős–Rényi random graphs, which represent an appropriate
infinite dimensional limit. Our SDRG algorithm is based on a judicious
elimination of most (irrelevant) new bonds generated under RG. It
yields excellent agreement with exact numerical results for universal
properties at the critical point without significant increase of computer
time, and confirm that, for Anderson localization, the upper critical
dimension $d_{uc}$ = infinite. We find excellent convergence of the
relevant $1/d$ expansion down to $d=2$, in contrast to the conventional
$2+\epsilon$ expansion, which has little to say about what happens
in any $d>3$. We show that the mysterious “mirror symmetry” of the
conductance scaling function is a genuine strong-coupling effect,
as speculated in early work \cite{vladPRL97MSR}. This opens an efficient
avenue to explore the critical properties of Anderson transition in
the strong-coupling limit in high dimensions.
\end{abstract}

\date{\today}

\pacs{71.10.Fd, 71.23.An, 71.30.+h, 72.15.Rn}

\maketitle
The Anderson transition is a nontrivial consequence of destructive
interference effects in disordered materials. Its simplest realization
is provided by the tight-binding model which describes electronic
states in a ``dirty'' conductor by mimicking the effect of impurities
through a random onsite potential. In spite of extensive studies,
one easily finds that some basic questions remain unanswered or in
disagreement. For instance, different values of the upper critical
dimension $d_{u}=4$, $6$, and $8$\ \cite{ThoulessPhysC1976,LukesPhysC1979,HarrisPRB1981,StarleyPRB1983,Suslov1996}
and $d_{u}=\infty$\ \cite{MirlinPRL1994,Peliti1986,GarciaPRB75}
have been reported. This question may look like purely academic but
indeed it has practical applications in quantum kicked rotor systems\ \cite{MoorPRL1995}
where the effective dimensionality of the dynamical localization is
determined by the number of incommensurate frequencies in the system.\ \cite{CasatiPRL1989}

One main challenge in investigating the localization transition is
the limited range of applicability of well known analytical approaches.
For example, The traditional “weak-localization” approach to the Anderson
metal-insulator transition is based on the fact that, in the vicinity
of the d=2, the transition is found at weak disorder, where perturbative
methods can be used; this lead to a flurry of results in 1980s. On
the other hand, more recent numerical results demonstrated that predictions
from such 2+epsilon expansions provide poor guidance even in d=3,
similarly as in other theories starting from the lower critical dimension.
\ \cite{wegner1980inverse,wegner1976,VollhardtPRL48,HeikoPRL76dimension,Kramer984D}
In contrast, the progress in numerical calculations during the last
20 years has increased dramatically our knowledge of the metal-insulator
transition, especially in dimensions such as $d=3$ and $4$ for which
a rigorous analytical treatment is not available. 

For most critical phenomena, the upper critical dimension has provided
a much better starting point, but so far such an approach has not
been available for Anderson localization. Since in high dimensions
the Anderson point shifts away from weak disorder, an appropriate
strong-disorder approach is called for. Here we show how an accurate
Strong Disorder Renormalization Group (SDRG) approach can be developed
for Anderson localization, where quantitatively accurate results can
be obtained in all dimensions. The SDRG method has been successful
in describing the critical and near-critical behavior of the Random
Transverse-Field Ising model and other random magnetic transitions,\ \cite{fisher92,igloi-review}
and have been recently used in electronic systems.\ \cite{GarelPRB80,HosseinJMPRB90}

Avoiding finite-size effects by having access to very large system
sizes and flexibility to work in any dimensions or topology in reasonable
time and computer memory resources have been always a long lived goal
for computational physicists. In this letter, we achieved this goal
by designing an efficient numerical approach able to study the localization
transition in the tight-binding model by computing the conductance
in all dimensions without much effort. Our implementation of the method
only keeps track of the main couplings in the system which allowed
us to greatly speed up the computer time.

\emph{Model and method.---} We study\textcolor{black}{{} the }$d$-dimensional
tight-binding model 

\begin{equation}
H=-\sum_{i,j}(t_{i,j}c_{i}^{\dagger}c_{j}^{\phantom{\dagger}}+\mathrm{h.c.})+\sum_{i}\varepsilon_{i}c_{i}^{\dagger}c_{i}^{\phantom{\dagger}},\label{eq:ham}
\end{equation}
where $c_{i}^{\dagger}$($c_{i}^{\phantom{\dagger}}$) is the canonical
creation (annihilation) operator of spineless fermions at site $i$,
$t_{i,j}=t_{j,i}$ is the hopping amplitude between sites $i$ and
$j$, and $\varepsilon_{i}$ is the onsite energy. The site energies
$\varepsilon_{i}$ are identically distributed random variables drawn
from a uniform distribution of zero mean and width $W$, and the hoping
amplitude $t_{i,j}=1$ if sites $i$ and $j$ are connected (which
is model dependent), otherwise it is zero. We treat this model using
the SDRG method\ \cite{HosseinJMPRB90} and compute the dimensionless
conductance defined as 

\begin{equation}
g\equiv g_{{\rm typ}}=\frac{\left\langle T\right\rangle _{{\rm geo}}}{1-\left\langle T\right\rangle _{{\rm geo}}}\label{eq:g-average}
\end{equation}
where $T$ is the transmittance, and $\left\langle \cdots\right\rangle _{{\rm geo}}$
denotes the geometric average. In this work, we will consider only
leads that are connected to single sites of the sample. Therefore,
$g$ is the two-point conductance.

The SDRG method consists in a iterative elimination of the strongest
energy scale $\Omega=\max\{|\varepsilon_{i}|,|t_{ij}|\}$ (with the
exception of those connected to the external wires) and renormalizing
the remaining ones couplings in the following fashion: (i) if $\Omega=|\varepsilon_{i}|$,
then site $i$ is eliminated from the system and the remaining couplings
are renormalized to 
\begin{equation}
\tilde{\varepsilon}_{k}=\varepsilon_{k}-\frac{t_{i,k}^{2}}{\varepsilon_{i}},\label{eq:newe-e}
\end{equation}
and 
\begin{equation}
\tilde{t}_{k,l}=t_{k,l}-\frac{t_{k,i}t_{i,l}}{\varepsilon_{i}};\label{eq:newt-e}
\end{equation}
on the other hand if (ii) $\Omega=|t_{i,j}|$, then sites $i$ and
$j$ are eliminated from the system yielding the renormalized couplings

\begin{equation}
\tilde{\varepsilon}_{k}=\varepsilon_{k}-\frac{\varepsilon_{i}t_{i,k}^{2}-2t_{i,j}t_{i,k}t_{j,k}+\varepsilon_{j}t_{j,k}^{2}}{t_{i,j}^{2}-\varepsilon_{i}\varepsilon_{j}},\label{eq:newe-t}
\end{equation}
 and 
\begin{equation}
\tilde{t}_{k,l}=t_{k,l}+\frac{\varepsilon_{j}t_{i,k}t_{i,l}-t_{i,j}(t_{i,k}t_{j,l}+t_{i,l}t_{j,k})+\varepsilon_{i}t_{j,k}t_{j,l}}{t_{i,j}^{2}-\varepsilon_{i}\varepsilon_{j}}.\label{eq:newt-t}
\end{equation}
In this way, we eliminate all the sites until there is a single renormalized
coupling $\tilde{\varepsilon}_{\alpha}$-$\tilde{t}_{\alpha,\beta}$-$\tilde{\varepsilon}_{\beta}$
connecting the leads at sites $\alpha$ and $\beta$ from which the
transmittance $T$ can be computed straightforwardly.

These transformations, although computed in perturbation theory, are
\emph{exact} in the purpose of studying transport properties (transmittance)
since it preserves the Green's function.\ \cite{aoki-80} As a consequence,
this method yields accurate results for the critical parameters associated
with the localization transition in any dimensions. However, as can
be seen from Eqs.\ (\ref{eq:newe-e})---(\ref{eq:newt-t}), the reconnection
of the lattice requires an increasing amount of memory and the procedure
becomes unpractical. In order to avoid this problem, many schemes
were proposed which are model dependent.\ \cite{motrunich-ising2d,hoyos-ladders,IgloiPRB83graph}
The modification of the SDRG scheme we adopt in this work is setting
a maximum coordination number $k_{{\rm max}}$ per site, i.e., we
follow the exact SDRG procedure but only keep track of the strongest
$k_{{\rm max}}$ couplings in each site. A detailed study comparing
the ``exact'' and ``modified'' SDRG procedures will be given elsewhere.\ \cite{unpublished}

\emph{Infinite dimensional limit.---}In Erdős–Rényi (ER) random graph,
we consider a system of $N\gg1$ sites in which two given sites $i$
and $j$ are connected with probability $p$ ($t_{i,j}=1$) and disconnected
with probability $1-p$ ($t_{i,j}=0$). Since the average number of
sites at a ``distance'' $L$ from a particular site increases exponentially
with $L$ , it effectively corresponds to the limit of $d\rightarrow\infty$.

In order to have a well defined length scale, the contact leads are
attached to two sites at the average shortest distance $L_{{\rm ER}}$,
where $L_{{\rm ER}}=\ln N/\ln\left\langle k\right\rangle $.\ \cite{Fronczak2004}
Here, $\left\langle k\right\rangle =p\left(N-1\right)$ is the average
coordination number which is chosen to be greater than the percolation
threshold $k_{c}=1$.\ \cite{Solebook} We verified that our final
results do not depend on the exact value of $\left\langle k\right\rangle $
as long as it is near and above $k_{c}$. 

\begin{figure}[h]
\begin{centering}
\includegraphics[clip,width=0.9\columnwidth]{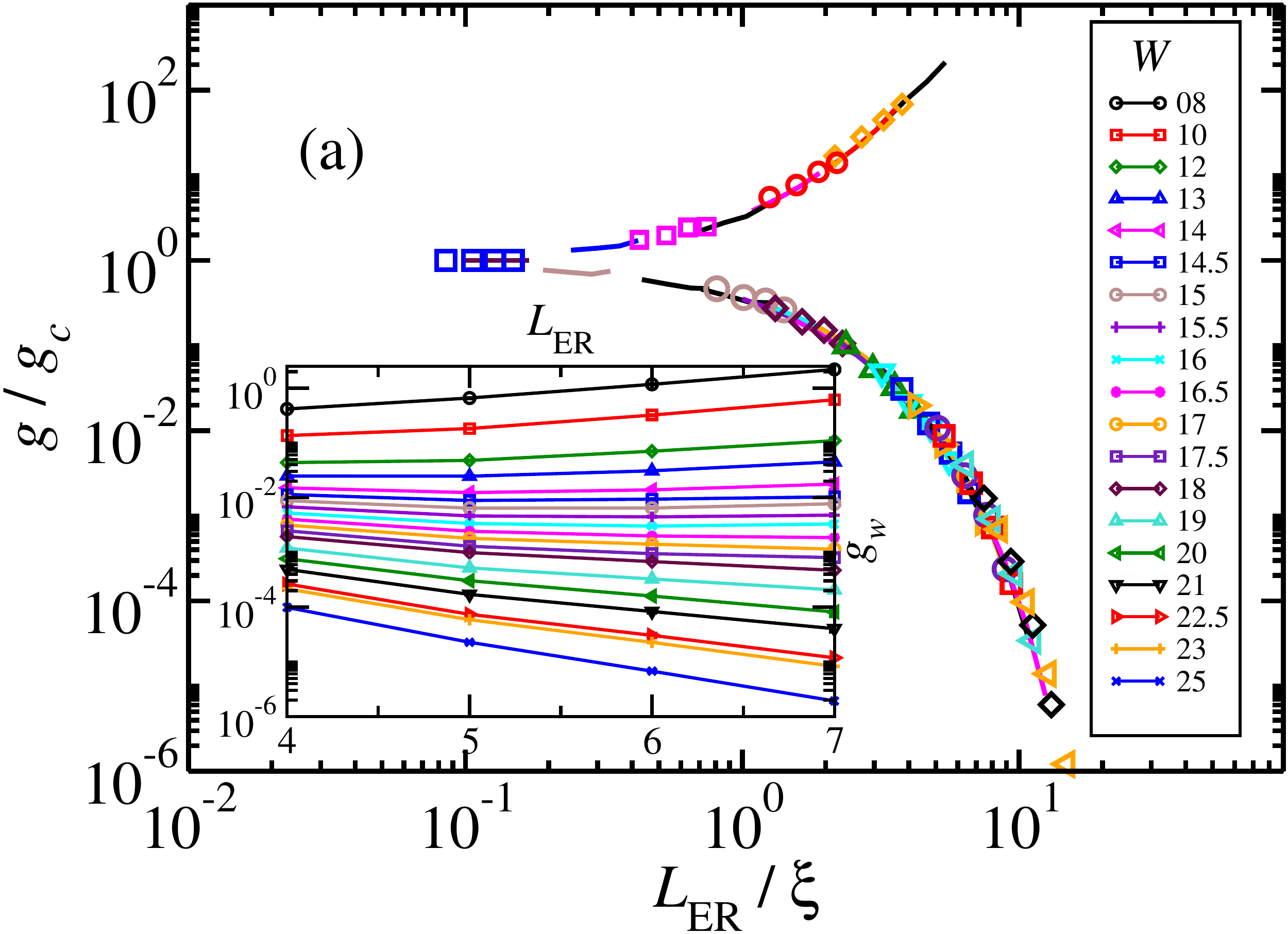}\\
\includegraphics[clip,width=0.9\columnwidth]{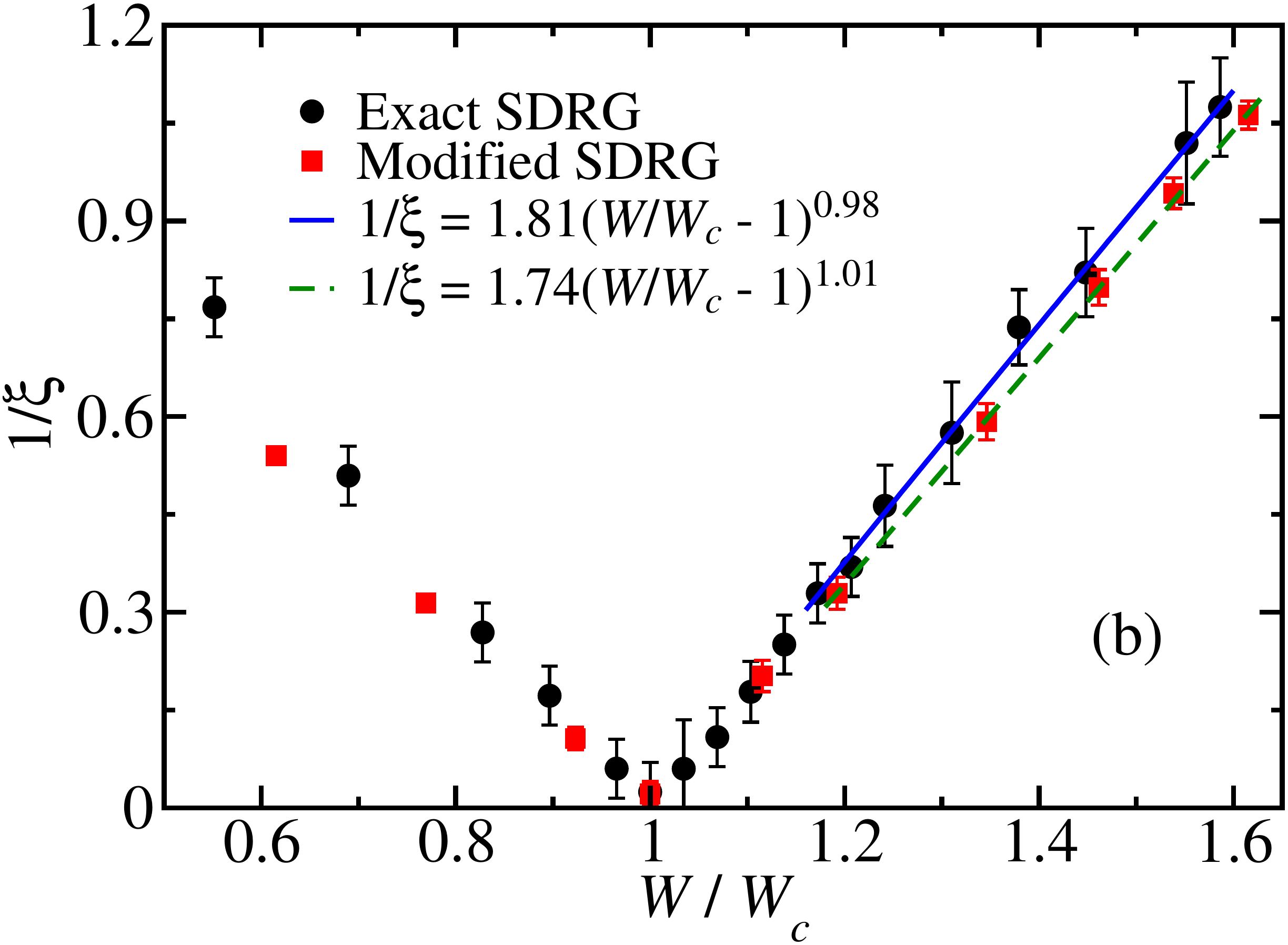}
\par\end{centering}

\protect\caption{(a) The typical two-point conductance $g$ of the ER random graph
with $\left\langle k\right\rangle =3.0$ and $N=3^{L_{{\rm ER}}}$,
$L_{{\rm ER}}=4,\dots,7$ for several disorder parameters $W=8$ up
to $W=25$ using the exact SDRG procedure (colorful solid lines) and
our modified algorithm (colorful symbols). We average over as many
disorder realizations needed to reach $5\%$ of precision (see main
text for details). Inset: the weighted conductance $g_{w}$ as a function
of $L_{{\rm ER}}$. Legends correspond to the inset, not the main
panel. (b) The localization length near the localization transition.
\label{fig:g-ER}}
\end{figure}

In similarity with previous studies on the Bethe lattice,\ \cite{Mirlin1991}
the distinction between the conducting and insulating phases manifests
in the different behavior of the ``weighted'' two-point conductance
$g_{w}(L)=N(L)g$. The extra factor $N(L)=\left\langle k\right\rangle (\left\langle k\right\rangle -1)^{L-1}$
counts the number of sites located at the distance $L$ from a given
site. It does not play any significant role in the universal behavior
of conductance {[}see Figs.\ \ref{fig:g-ER}(a) and (b){]} but is
useful to pinpoint the critical point $W_{c}$ {[}see inset of Fig.\ \ref{fig:g-ER}(a){]}
and to obtain the localization length $\xi$ in the localized phase:
$\ln g_{w}\sim-L/\xi$. The critical disorder value $W_{c}=14.5(3)$
(exact SDRG) and $W_{c}=13.0(3)$ (modified SDRG with $k_{{\rm max}}=20$).
Our estimate for the localization length exponent (defined via $\xi\sim|W-W_{c}|^{-\nu}$
considering only $\xi$ that are less than $L_{{\rm ER}}$) is $\nu=0.98(4)$
(exact SDRG) and $\nu=1.01(5)$ (modified SDRG) which is very close
to the exact value $\nu=1$ in $d\rightarrow\infty$.\ \cite{GarelCayleyTree2009,Mirlin1991}
In the metallic phase, the localization length is obtained by dividing
$L$ by $\xi$ such that all the curves $g_{w}/g_{wc}$ collapse in
a single curve. This procedure is precise up to an irrelevant global
pre-factor. In this way, we confirm that $\nu$ is the same in both
localized and delocalized phases (within the statistical error).

\begin{figure}[h]
\begin{centering}
\includegraphics[clip,width=0.9\columnwidth]{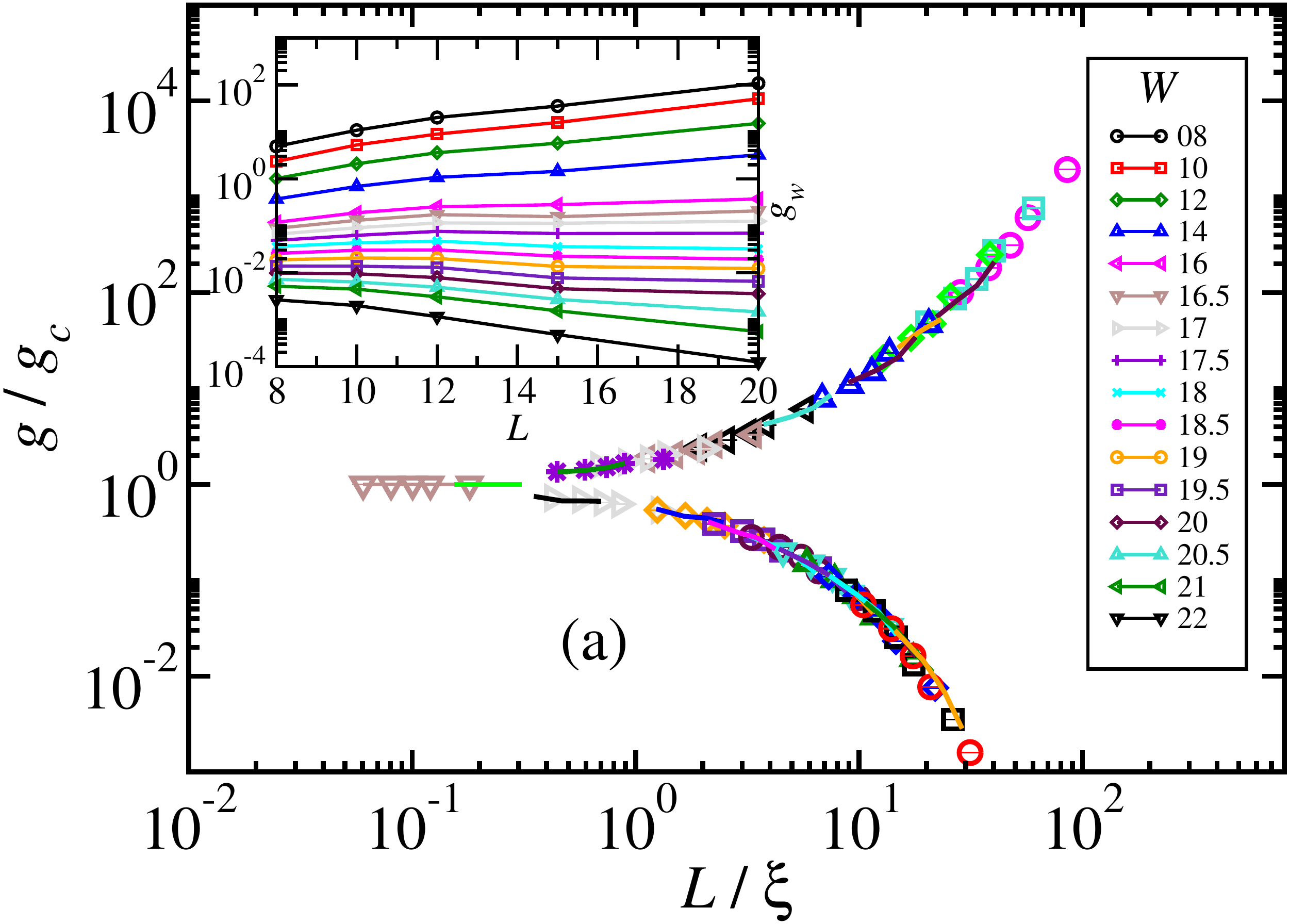}\\
\includegraphics[clip,width=0.9\columnwidth]{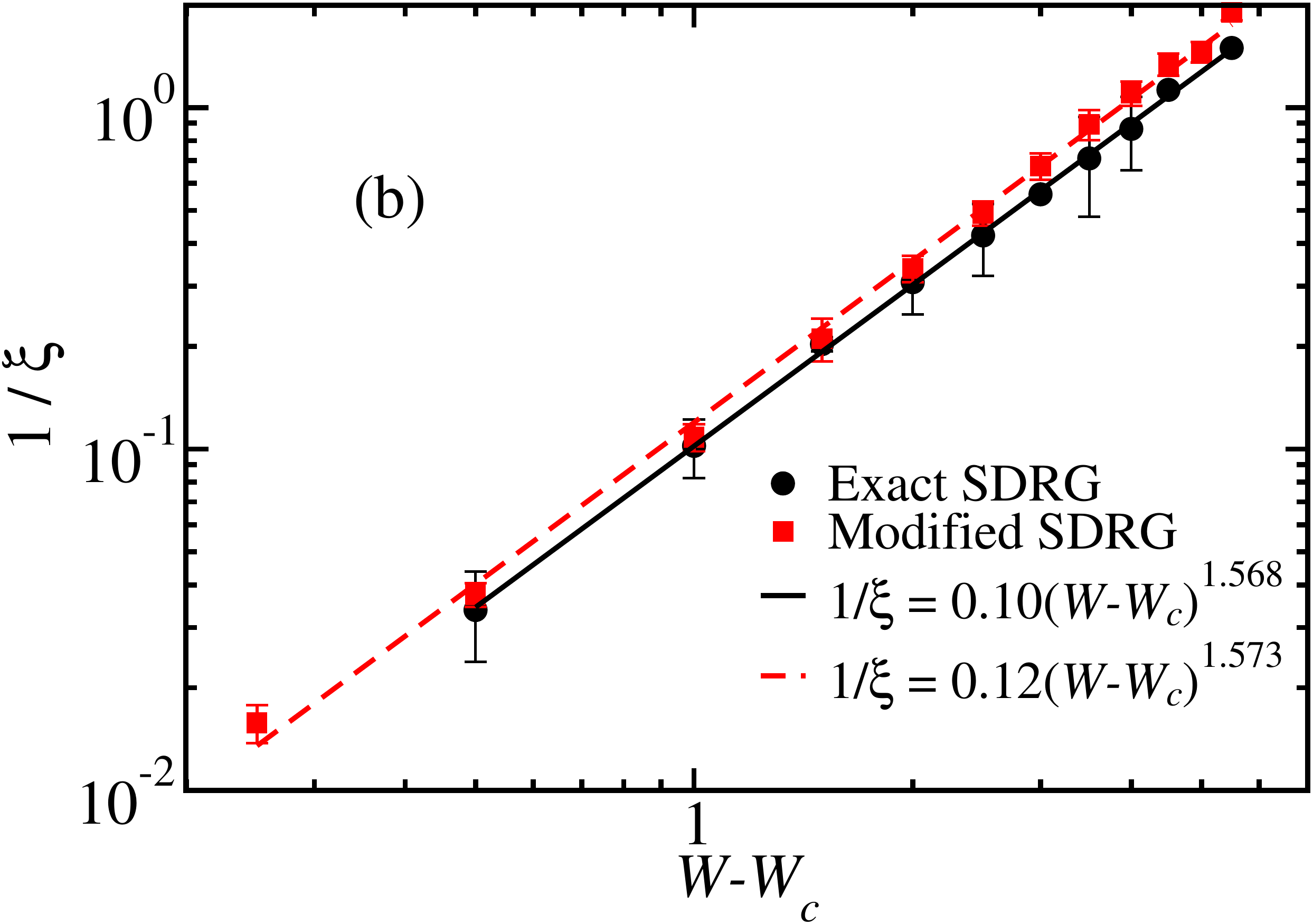}
\par\end{centering}

\protect\caption{(a) The typical two-point conductance $g$ for the 3D cubic lattice
for various disorder parameters using the exact SDRG procedure (colorful
solid lines) and our proposed modified algorithm (colorful symbols).
Inset: the weighted two-point conductance $g_{w}$ obtained from the
exact SDRG method. Legends correspond to the inset, not the main panel.
(b) The localization length $\xi$ (in the localized phase) as a function
of the distance from criticality $W-W_{c}$. \label{fig:g-3D}}
\end{figure}

\emph{Cubic lattice in }$d=3$\emph{.---}We now apply the SDRG method
to the cubic lattice in $d=3$. Here, $t_{i,j}=1$ if $i$ and $j$
are nearest neighbors, and $t_{i,j}=0$ otherwise. The value of upper
cutoff $k_{{\rm max}}$ can be adjusted according to desired accuracy.
Here, as in the ER graph, a modest value of $k_{{\rm max}}=20$ is
sufficient for getting good agreement between the exact and modified
SDRG methods (within the $5\%$ of the statistical accuracy). We have
used chains of sizes $L=8,$ $10$, $12$, $15$, and $20$ with periodic
boundary conditions and the leads were attached to the corner and
to the center sites of the sample (maximum possible distance). In
the inset of Fig.\ \ref{fig:g-3D}(a), we plot $g_{w}=L^{3}g$ for
various disorder parameter $W$. Unlike the ER graph, it is not so
simple to pinpoint the critical point $W_{c}$, mainly because $g_{w}$
at criticality is not constant for large $L$. We then try scaling
using different critical $W_{c}$ until the best data collapse is
obtained {[}see Fig.\ \ref{fig:g-3D}(a){]}. We find $W_{c}=16.5(5)$
(exact SDRG) and $W_{c}=17.5(5)$ (modified SDRG). The localization
length exponent is obtained in the same way as in the ER graph {[}see
Fig.\ \ref{fig:g-3D}(b){]} from which we obtained $\nu=1.57(1)$
in agreement with previous results.\ \cite{slevin-ohtsuki-prl99,rodriguez-etal-prb11}
Although this result is obtained by fitting only those data in which
$\xi<20$, it fits quite well all the entire data set.

Figure 3 presents a study of dimensional dependency of $\nu$ at $d=3,\ 4,\ 6,\ 10,$
and infinity using different approaches. It manifests the limited
range of applicability of some well-known analytical theories such
as $2+\epsilon$ expansion, the self-consistent theory proposed by
D. Vollhardt and P. Wo\"{l}fle, and phenomenological proposal of the
beta function in $d\geq1$ by Shapiro's work. They lead to very poor
results for $\nu$ as $d$ becomes equal or higher than $3$. Our
modified SDRG algorithm estimations for $\nu$ are consistent with
the most recent numerical computation done by Y. Ueoka and K. Slevin
on dimensions up to $5$. \cite{Slevin2014} It is clear from our
data, the upper critical dimension is not at finite dimensions predicted
in references \cite{ThoulessPhysC1976,LukesPhysC1979,HarrisPRB1981,StarleyPRB1983,Suslov1996}while
infinite dimension seems more reasonable candidate for the upper critical
dimension of Anderson model. In contrast to García-García work, we
find $1/\nu$ follows a behavior more complicated than a linear relationship
with $1/d$ (a cubic function fits well with our data points in Fig.
3) . Our modified algorithm allows us to study higher dimensions without
limiting us to very small system sizes which provides us a better
prediction for the behavior described in Fig. 3. 

\begin{figure}[h]
\includegraphics[width=1\columnwidth]{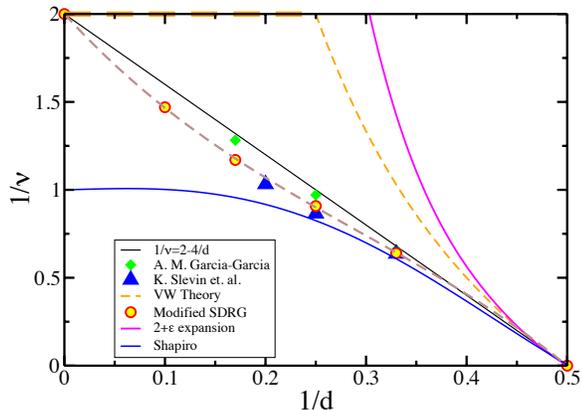}

\protect\caption{The inverse of critical exponent $\nu$ is plotted versus $1/d$ for
$d\geq2$ , along with some recent numerical estimations of this exponent
in higher dimensions by Slevin et. al. \cite{Slevin2014}(Blue triangles)
and García-García \cite{GarciaPRB75}(Green diamonds). Analytical
predictions of three well-known theories are included to stress the
limited range of reliability of them to estimate $\nu$ as it is discussed
in the main text. The Brown color dashed line is a cubic function
fitted to our data presenting the number of orders of $1/d$ needed
to describe the localization transition for $d\geq2$ . The coefficients
of cubic fitted function are computed as $c_{0}=2.000\pm0.008,\ c_{1}=-6.263\pm0.167,\ c_{2}=10.380\pm0.864,$
and $c_{3}=-11.713\pm1.1458$, respectively.}
\end{figure}

In contrast to weak disorder approach, our approach based on strong
disorder limit presents much more reasonable results in estimating
the critical exponent for $d\geq3$. The significant role of strong
coupling in localization transition is also prominent in study of
the mirror symmetry phenomena which has been also observed experimentally.
\cite{HsuVallesPRL95,kravchenkoPRL96}The Mirror Symmetry Range (MSR)
can be defined as the range of $g/g_{c}$ where mirror symmetry in
the scaling function holds. The mirror symmetry idea was proposed
in early work for two dimensional MIT, where they conclude that the
related experimental results provide striking evidence about the form
of the beta function in the critical region. \cite{vladPRL97MSR}
In particular, they indicate that in a wide range of conductances
the beta function is well-approximated by the linear expression in
$t=\ln(g/g_{c})$. This observation can be interpreted by noticing
that deep in the insulating regime, $(g<<g_{c})$ the beta function
is exactly given by $\beta(g)\sim\ln(g)$. The related experimental
result can thus be interpreted as evidence that the same slow logarithmic
form of the beta function persists beyond the insulating limit well
into the critical regime. This feature could be used as a basis of
approximate calculations of the critical exponents. In contrast to
the well-known $2+\epsilon$ expansion, here one would try to obtain
the form of the beta function in the critical region by an expansion
around the strong disorder limit.

\begin{figure}[h]
\includegraphics[width=1\columnwidth]{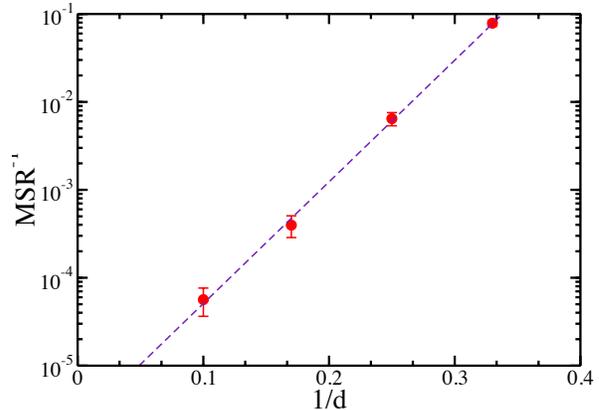}

\protect\caption{The inverse of Mirror Symmetry Range (MSR) as a function of $1/d$
presents an increasing trend implying the strong-coupling effect.
The dashed line presents the fitting function expressed as $MSR^{-1}=a\exp(b/d)$
where $a=2.10\times10^{-6}$ and $b=31.86\pm0.88$. }
\end{figure}

\emph{Applications.---}Since the number of sites increases rapidly
with lattice size in higher dimensions, depending on accessible computer
memory, one might be limited to try arbitrary large system size. In
this situation, an alternative solution is to use a site percolated
hyper-cubic lattice instead, to decrease the number of sites which
actually play important role in the transition (We expect the critical
behavior does not vary with $<k>$ as long as it is above its classical
percolation threshold.). Another possible application of our method
is to consider the global conductance instead of the two-point one.
In this case, the leads will be attached to opposite planes of the
sample which will be fully connected after the interior is decimated
out.

\emph{Conclusion.---}We showed our modified SDRG method is suitable
to study the localization properties of large system in any dimension.
This significant progress eliminates previous obstacles such as computational
time or computer memory size and paves the avenue for future study
of the Anderson transition.\emph{ }We provided concluding evidence
that the upper critical dimension for localization is infinity. Considering
neither the self-consistent theory of localization exact for the Cayley
tree nor the $\epsilon$-expansion formalism is accurate for intermediate
dimensions, we proposed a strong coupling basis for the localization
problem and infinite dimension as starting point.
\begin{acknowledgments}
This work was supported by the NSF grants DMR-1005751, DMR-1410132
and PHYS-1066293, by the National High Magnetic Field Laboratory,
by the Simons Foundation, by FAPESP under Grants 07/57630-5 and 2013/09850-7,
and by CNPq under Grants 304311/2010-3, 590093/2011-8 and 305261/2012-6.
We acknowledge the hospitality of the Aspen Center for Physics.
\end{acknowledgments}

\bibliographystyle{apsrev4-1}
\bibliography{bibliography}

\begin{thebibliography}{34}%
\makeatletter
\providecommand \@ifxundefined [1]{%
 \@ifx{#1\undefined}
}%
\providecommand \@ifnum [1]{%
 \ifnum #1\expandafter \@firstoftwo
 \else \expandafter \@secondoftwo
 \fi
}%
\providecommand \@ifx [1]{%
 \ifx #1\expandafter \@firstoftwo
 \else \expandafter \@secondoftwo
 \fi
}%
\providecommand \natexlab [1]{#1}%
\providecommand \enquote  [1]{``#1''}%
\providecommand \bibnamefont  [1]{#1}%
\providecommand \bibfnamefont [1]{#1}%
\providecommand \citenamefont [1]{#1}%
\providecommand \href@noop [0]{\@secondoftwo}%
\providecommand \href [0]{\begingroup \@sanitize@url \@href}%
\providecommand \@href[1]{\@@startlink{#1}\@@href}%
\providecommand \@@href[1]{\endgroup#1\@@endlink}%
\providecommand \@sanitize@url [0]{\catcode `\\12\catcode `\$12\catcode
  `\&12\catcode `\#12\catcode `\^12\catcode `\_12\catcode `\%12\relax}%
\providecommand \@@startlink[1]{}%
\providecommand \@@endlink[0]{}%
\providecommand \url  [0]{\begingroup\@sanitize@url \@url }%
\providecommand \@url [1]{\endgroup\@href {#1}{\urlprefix }}%
\providecommand \urlprefix  [0]{URL }%
\providecommand \Eprint [0]{\href }%
\providecommand \doibase [0]{http://dx.doi.org/}%
\providecommand \selectlanguage [0]{\@gobble}%
\providecommand \bibinfo  [0]{\@secondoftwo}%
\providecommand \bibfield  [0]{\@secondoftwo}%
\providecommand \translation [1]{[#1]}%
\providecommand \BibitemOpen [0]{}%
\providecommand \bibitemStop [0]{}%
\providecommand \bibitemNoStop [0]{.\EOS\space}%
\providecommand \EOS [0]{\spacefactor3000\relax}%
\providecommand \BibitemShut  [1]{\csname bibitem#1\endcsname}%
\let\auto@bib@innerbib\@empty
\bibitem [{\citenamefont {Dobrosavljevi\ifmmode~\acute{c}\else \'{c}\fi{}}\
  \emph {et~al.}(1997)\citenamefont {Dobrosavljevi\ifmmode~\acute{c}\else
  \'{c}\fi{}}, \citenamefont {Abrahams}, \citenamefont {Miranda},\ and\
  \citenamefont {Chakravarty}}]{vladPRL97MSR}%
  \BibitemOpen
  \bibfield  {author} {\bibinfo {author} {\bibfnamefont {V.}~\bibnamefont
  {Dobrosavljevi\ifmmode~\acute{c}\else \'{c}\fi{}}}, \bibinfo {author}
  {\bibfnamefont {E.}~\bibnamefont {Abrahams}}, \bibinfo {author}
  {\bibfnamefont {E.}~\bibnamefont {Miranda}}, \ and\ \bibinfo {author}
  {\bibfnamefont {S.}~\bibnamefont {Chakravarty}},\ }\href@noop {} {\bibfield
  {journal} {\bibinfo  {journal} {Phys. Rev. Lett.}\ }\textbf {\bibinfo
  {volume} {79}},\ \bibinfo {pages} {455} (\bibinfo {year} {1997})}\BibitemShut
  {NoStop}%
\bibitem [{\citenamefont {Thouless}(1976)}]{ThoulessPhysC1976}%
  \BibitemOpen
  \bibfield  {author} {\bibinfo {author} {\bibfnamefont {D.~J.}\ \bibnamefont
  {Thouless}},\ }\href {http://stacks.iop.org/0022-3719/9/i=21/a=003}
  {\bibfield  {journal} {\bibinfo  {journal} {Journal of Physics C: Solid State
  Physics}\ }\textbf {\bibinfo {volume} {9}},\ \bibinfo {pages} {L603}
  (\bibinfo {year} {1976})}\BibitemShut {NoStop}%
\bibitem [{\citenamefont {Lukes}(1979)}]{LukesPhysC1979}%
  \BibitemOpen
  \bibfield  {author} {\bibinfo {author} {\bibfnamefont {T.}~\bibnamefont
  {Lukes}},\ }\href {http://stacks.iop.org/0022-3719/12/i=20/a=006} {\bibfield
  {journal} {\bibinfo  {journal} {Journal of Physics C: Solid State Physics}\
  }\textbf {\bibinfo {volume} {12}},\ \bibinfo {pages} {L797} (\bibinfo {year}
  {1979})}\BibitemShut {NoStop}%
\bibitem [{\citenamefont {Harris}\ and\ \citenamefont
  {Lubensky}(1981)}]{HarrisPRB1981}%
  \BibitemOpen
  \bibfield  {author} {\bibinfo {author} {\bibfnamefont {A.}~\bibnamefont
  {Harris}}\ and\ \bibinfo {author} {\bibfnamefont {T.}~\bibnamefont
  {Lubensky}},\ }\href {\doibase 10.1103/PhysRevB.23.2640} {\bibfield
  {journal} {\bibinfo  {journal} {Phys. Rev. B}\ }\textbf {\bibinfo {volume}
  {23}},\ \bibinfo {pages} {2640} (\bibinfo {year} {1981})}\BibitemShut
  {NoStop}%
\bibitem [{\citenamefont {Straley}(1983)}]{StarleyPRB1983}%
  \BibitemOpen
  \bibfield  {author} {\bibinfo {author} {\bibfnamefont {J.}~\bibnamefont
  {Straley}},\ }\href {\doibase 10.1103/PhysRevB.28.5393} {\bibfield  {journal}
  {\bibinfo  {journal} {Phys. Rev. B}\ }\textbf {\bibinfo {volume} {28}},\
  \bibinfo {pages} {5393} (\bibinfo {year} {1983})}\BibitemShut {NoStop}%
\bibitem [{\citenamefont {Suslov}(1996)}]{Suslov1996}%
  \BibitemOpen
  \bibfield  {author} {\bibinfo {author} {\bibfnamefont {I.}~\bibnamefont
  {Suslov}},\ }\href {\doibase 10.1134/1.567110} {\bibfield  {journal}
  {\bibinfo  {journal} {Journal of Experimental and Theoretical Physics
  Letters}\ }\textbf {\bibinfo {volume} {63}},\ \bibinfo {pages} {895}
  (\bibinfo {year} {1996})}\BibitemShut {NoStop}%
\bibitem [{\citenamefont {Mirlin}\ and\ \citenamefont
  {Fyodorov}(1994)}]{MirlinPRL1994}%
  \BibitemOpen
  \bibfield  {author} {\bibinfo {author} {\bibfnamefont {A.}~\bibnamefont
  {Mirlin}}\ and\ \bibinfo {author} {\bibfnamefont {Y.}~\bibnamefont
  {Fyodorov}},\ }\href {\doibase 10.1103/PhysRevLett.72.526} {\bibfield
  {journal} {\bibinfo  {journal} {Phys. Rev. Lett.}\ }\textbf {\bibinfo
  {volume} {72}},\ \bibinfo {pages} {526} (\bibinfo {year} {1994})}\BibitemShut
  {NoStop}%
\bibitem [{\citenamefont {Castellani}\ \emph {et~al.}(1986)\citenamefont
  {Castellani}, \citenamefont {Castro},\ and\ \citenamefont
  {Peliti}}]{Peliti1986}%
  \BibitemOpen
  \bibfield  {author} {\bibinfo {author} {\bibfnamefont {C.}~\bibnamefont
  {Castellani}}, \bibinfo {author} {\bibfnamefont {C.~D.}\ \bibnamefont
  {Castro}}, \ and\ \bibinfo {author} {\bibfnamefont {L.}~\bibnamefont
  {Peliti}},\ }\href {http://stacks.iop.org/0305-4470/19/i=17/a=009} {\bibfield
   {journal} {\bibinfo  {journal} {Journal of Physics A: Mathematical and
  General}\ }\textbf {\bibinfo {volume} {19}},\ \bibinfo {pages} {L1099}
  (\bibinfo {year} {1986})}\BibitemShut {NoStop}%
\bibitem [{\citenamefont {Garc{\'\i}a-Garc{\'\i}a}\ and\ \citenamefont
  {Cuevas}(2007)}]{GarciaPRB75}%
  \BibitemOpen
  \bibfield  {author} {\bibinfo {author} {\bibfnamefont {A.~M.}\ \bibnamefont
  {Garc{\'\i}a-Garc{\'\i}a}}\ and\ \bibinfo {author} {\bibfnamefont
  {E.}~\bibnamefont {Cuevas}},\ }\href {\doibase 10.1103/PhysRevB.75.174203}
  {\bibfield  {journal} {\bibinfo  {journal} {Phys. Rev. B}\ }\textbf {\bibinfo
  {volume} {75}},\ \bibinfo {pages} {174203} (\bibinfo {year}
  {2007})}\BibitemShut {NoStop}%
\bibitem [{\citenamefont {Moore}\ \emph {et~al.}(1995)\citenamefont {Moore},
  \citenamefont {Robinson}, \citenamefont {Bharucha}, \citenamefont
  {Sundaram},\ and\ \citenamefont {Raizen}}]{MoorPRL1995}%
  \BibitemOpen
  \bibfield  {author} {\bibinfo {author} {\bibfnamefont {F.}~\bibnamefont
  {Moore}}, \bibinfo {author} {\bibfnamefont {J.}~\bibnamefont {Robinson}},
  \bibinfo {author} {\bibfnamefont {C.}~\bibnamefont {Bharucha}}, \bibinfo
  {author} {\bibfnamefont {B.}~\bibnamefont {Sundaram}}, \ and\ \bibinfo
  {author} {\bibfnamefont {M.}~\bibnamefont {Raizen}},\ }\href {\doibase
  10.1103/PhysRevLett.75.4598} {\bibfield  {journal} {\bibinfo  {journal}
  {Phys. Rev. Lett.}\ }\textbf {\bibinfo {volume} {75}},\ \bibinfo {pages}
  {4598} (\bibinfo {year} {1995})}\BibitemShut {NoStop}%
\bibitem [{\citenamefont {Casati}\ \emph {et~al.}(1989)\citenamefont {Casati},
  \citenamefont {Guarneri},\ and\ \citenamefont
  {Shepelyansky}}]{CasatiPRL1989}%
  \BibitemOpen
  \bibfield  {author} {\bibinfo {author} {\bibfnamefont {G.}~\bibnamefont
  {Casati}}, \bibinfo {author} {\bibfnamefont {I.}~\bibnamefont {Guarneri}}, \
  and\ \bibinfo {author} {\bibfnamefont {D.}~\bibnamefont {Shepelyansky}},\
  }\href {\doibase 10.1103/PhysRevLett.62.345} {\bibfield  {journal} {\bibinfo
  {journal} {Phys. Rev. Lett.}\ }\textbf {\bibinfo {volume} {62}},\ \bibinfo
  {pages} {345} (\bibinfo {year} {1989})}\BibitemShut {NoStop}%
\bibitem [{\citenamefont {Wegner}(1980)}]{wegner1980inverse}%
  \BibitemOpen
  \bibfield  {author} {\bibinfo {author} {\bibfnamefont {F.}~\bibnamefont
  {Wegner}},\ }\href@noop {} {\bibfield  {journal} {\bibinfo  {journal}
  {Zeitschrift f{\"u}r Physik B Condensed Matter}\ }\textbf {\bibinfo {volume}
  {36}},\ \bibinfo {pages} {209} (\bibinfo {year} {1980})}\BibitemShut
  {NoStop}%
\bibitem [{\citenamefont {Wegner}(1976)}]{wegner1976}%
  \BibitemOpen
  \bibfield  {author} {\bibinfo {author} {\bibfnamefont {F.}~\bibnamefont
  {Wegner}},\ }\href {\doibase 10.1007/BF01315248} {\bibfield  {journal}
  {\bibinfo  {journal} {Zeitschrift f{\"u}r Physik B Condensed Matter}\
  }\textbf {\bibinfo {volume} {25}},\ \bibinfo {pages} {327} (\bibinfo {year}
  {1976})}\BibitemShut {NoStop}%
\bibitem [{\citenamefont {Vollhardt}\ and\ \citenamefont
  {W\"olfle}(1982)}]{VollhardtPRL48}%
  \BibitemOpen
  \bibfield  {author} {\bibinfo {author} {\bibfnamefont {D.}~\bibnamefont
  {Vollhardt}}\ and\ \bibinfo {author} {\bibfnamefont {P.}~\bibnamefont
  {W\"olfle}},\ }\href {\doibase 10.1103/PhysRevLett.48.699} {\bibfield
  {journal} {\bibinfo  {journal} {Phys. Rev. Lett.}\ }\textbf {\bibinfo
  {volume} {48}},\ \bibinfo {pages} {699} (\bibinfo {year} {1982})}\BibitemShut
  {NoStop}%
\bibitem [{\citenamefont {Schreiber}\ and\ \citenamefont
  {Grussbach}(1996)}]{HeikoPRL76dimension}%
  \BibitemOpen
  \bibfield  {author} {\bibinfo {author} {\bibfnamefont {M.}~\bibnamefont
  {Schreiber}}\ and\ \bibinfo {author} {\bibfnamefont {H.}~\bibnamefont
  {Grussbach}},\ }\href {\doibase 10.1103/PhysRevLett.76.1687} {\bibfield
  {journal} {\bibinfo  {journal} {Phys. Rev. Lett.}\ }\textbf {\bibinfo
  {volume} {76}},\ \bibinfo {pages} {1687} (\bibinfo {year}
  {1996})}\BibitemShut {NoStop}%
\bibitem [{\citenamefont {I.~Kh.~Zharekeshev}(1998)}]{Kramer984D}%
  \BibitemOpen
  \bibfield  {author} {\bibinfo {author} {\bibfnamefont {B.~K.}\ \bibnamefont
  {I.~Kh.~Zharekeshev}},\ }\href
  {http://link.aps.org/doi/10.1103/PhysRevLett.76.1687} {\bibfield  {journal}
  {\bibinfo  {journal} {Ann. Phys. (Leipzig)}\ }\textbf {\bibinfo {volume}
  {7}},\ \bibinfo {pages} {442} (\bibinfo {year} {1998})}\BibitemShut {NoStop}%
\bibitem [{\citenamefont {Fisher}(1992)}]{fisher92}%
  \BibitemOpen
  \bibfield  {author} {\bibinfo {author} {\bibfnamefont {D.~S.}\ \bibnamefont
  {Fisher}},\ }\href {\doibase 10.1103/PhysRevLett.69.534} {\bibfield
  {journal} {\bibinfo  {journal} {Phys. Rev. Lett.}\ }\textbf {\bibinfo
  {volume} {69}},\ \bibinfo {pages} {534} (\bibinfo {year} {1992})}\BibitemShut
  {NoStop}%
\bibitem [{\citenamefont {Igl\'oi}\ and\ \citenamefont
  {Monthus}(2005)}]{igloi-review}%
  \BibitemOpen
  \bibfield  {author} {\bibinfo {author} {\bibfnamefont {F.}~\bibnamefont
  {Igl\'oi}}\ and\ \bibinfo {author} {\bibfnamefont {C.}~\bibnamefont
  {Monthus}},\ }\href {\doibase 10.1016/j.physrep.2005.02.006} {\bibfield
  {journal} {\bibinfo  {journal} {Phys. Rep.}\ }\textbf {\bibinfo {volume}
  {412}},\ \bibinfo {pages} {277} (\bibinfo {year} {2005})}\BibitemShut
  {NoStop}%
\bibitem [{\citenamefont {Monthus}\ and\ \citenamefont
  {Garel}(2009{\natexlab{a}})}]{GarelPRB80}%
  \BibitemOpen
  \bibfield  {author} {\bibinfo {author} {\bibfnamefont {C.}~\bibnamefont
  {Monthus}}\ and\ \bibinfo {author} {\bibfnamefont {T.}~\bibnamefont
  {Garel}},\ }\href {\doibase 10.1103/PhysRevB.80.024203} {\bibfield  {journal}
  {\bibinfo  {journal} {Phys. Rev. B}\ }\textbf {\bibinfo {volume} {80}},\
  \bibinfo {pages} {024203} (\bibinfo {year} {2009}{\natexlab{a}})}\BibitemShut
  {NoStop}%
\bibitem [{\citenamefont {Javan~Mard}\ \emph {et~al.}(2014)\citenamefont
  {Javan~Mard}, \citenamefont {Hoyos}, \citenamefont {Miranda},\ and\
  \citenamefont {Dobrosavljevi\ifmmode~\acute{c}\else
  \'{c}\fi{}}}]{HosseinJMPRB90}%
  \BibitemOpen
  \bibfield  {author} {\bibinfo {author} {\bibfnamefont {H.}~\bibnamefont
  {Javan~Mard}}, \bibinfo {author} {\bibfnamefont {J.~A.}\ \bibnamefont
  {Hoyos}}, \bibinfo {author} {\bibfnamefont {E.}~\bibnamefont {Miranda}}, \
  and\ \bibinfo {author} {\bibfnamefont {V.}~\bibnamefont
  {Dobrosavljevi\ifmmode~\acute{c}\else \'{c}\fi{}}},\ }\href {\doibase
  10.1103/PhysRevB.90.125141} {\bibfield  {journal} {\bibinfo  {journal} {Phys.
  Rev. B}\ }\textbf {\bibinfo {volume} {90}},\ \bibinfo {pages} {125141}
  (\bibinfo {year} {2014})}\BibitemShut {NoStop}%
\bibitem [{\citenamefont {Aoki}(1980)}]{aoki-80}%
  \BibitemOpen
  \bibfield  {author} {\bibinfo {author} {\bibfnamefont {H.}~\bibnamefont
  {Aoki}},\ }\href {http://stacks.iop.org/0022-3719/13/i=18/a=006} {\bibfield
  {journal} {\bibinfo  {journal} {Journal of Physics C: Solid State Physics}\
  }\textbf {\bibinfo {volume} {13}},\ \bibinfo {pages} {3369} (\bibinfo {year}
  {1980})}\BibitemShut {NoStop}%
\bibitem [{\citenamefont {Motrunich}\ \emph {et~al.}(2000)\citenamefont
  {Motrunich}, \citenamefont {Mau}, \citenamefont {Huse},\ and\ \citenamefont
  {Fisher}}]{motrunich-ising2d}%
  \BibitemOpen
  \bibfield  {author} {\bibinfo {author} {\bibfnamefont {O.}~\bibnamefont
  {Motrunich}}, \bibinfo {author} {\bibfnamefont {S.-C.}\ \bibnamefont {Mau}},
  \bibinfo {author} {\bibfnamefont {D.~A.}\ \bibnamefont {Huse}}, \ and\
  \bibinfo {author} {\bibfnamefont {D.~S.}\ \bibnamefont {Fisher}},\ }\href
  {\doibase 10.1103/PhysRevB.61.1160} {\bibfield  {journal} {\bibinfo
  {journal} {Phys. Rev. B}\ }\textbf {\bibinfo {volume} {61}},\ \bibinfo
  {pages} {1160} (\bibinfo {year} {2000})}\BibitemShut {NoStop}%
\bibitem [{\citenamefont {Hoyos}\ and\ \citenamefont
  {Miranda}(2004)}]{hoyos-ladders}%
  \BibitemOpen
  \bibfield  {author} {\bibinfo {author} {\bibfnamefont {J.~A.}\ \bibnamefont
  {Hoyos}}\ and\ \bibinfo {author} {\bibfnamefont {E.}~\bibnamefont
  {Miranda}},\ }\href {\doibase 10.1103/PhysRevB.69.214411} {\bibfield
  {journal} {\bibinfo  {journal} {Phys. Rev. B}\ }\textbf {\bibinfo {volume}
  {69}},\ \bibinfo {pages} {214411} (\bibinfo {year} {2004})}\BibitemShut
  {NoStop}%
\bibitem [{\citenamefont {Kov\'acs}\ and\ \citenamefont
  {Igl\'oi}(2011)}]{IgloiPRB83graph}%
  \BibitemOpen
  \bibfield  {author} {\bibinfo {author} {\bibfnamefont {I.~A.}\ \bibnamefont
  {Kov\'acs}}\ and\ \bibinfo {author} {\bibfnamefont {F.}~\bibnamefont
  {Igl\'oi}},\ }\href {\doibase 10.1103/PhysRevB.83.174207} {\bibfield
  {journal} {\bibinfo  {journal} {Phys. Rev. B}\ }\textbf {\bibinfo {volume}
  {83}},\ \bibinfo {pages} {174207} (\bibinfo {year} {2011})}\BibitemShut
  {NoStop}%
\bibitem [{\citenamefont {{H. Javan Mard, J. A. Hoyos, E. Miranda, and V.
  Dobrosavljevi\'c}}()}]{unpublished}%
  \BibitemOpen
  \bibfield  {author} {\bibinfo {author} {\bibnamefont {{H. Javan Mard, J. A.
  Hoyos, E. Miranda, and V. Dobrosavljevi\'c}}},\ }\href@noop {} {}\bibinfo
  {note} {Unpublished}\BibitemShut {NoStop}%
\bibitem [{\citenamefont {Fronczak}\ \emph {et~al.}(2004)\citenamefont
  {Fronczak}, \citenamefont {Fronczak},\ and\ \citenamefont
  {Ho\l{}yst}}]{Fronczak2004}%
  \BibitemOpen
  \bibfield  {author} {\bibinfo {author} {\bibfnamefont {A.}~\bibnamefont
  {Fronczak}}, \bibinfo {author} {\bibfnamefont {P.}~\bibnamefont {Fronczak}},
  \ and\ \bibinfo {author} {\bibfnamefont {J.~A.}\ \bibnamefont {Ho\l{}yst}},\
  }\href@noop {} {\bibfield  {journal} {\bibinfo  {journal} {Phys. Rev. E}\
  }\textbf {\bibinfo {volume} {70}},\ \bibinfo {pages} {56110} (\bibinfo {year}
  {2004})}\BibitemShut {NoStop}%
\bibitem [{\citenamefont {Sole}(2011)}]{Solebook}%
  \BibitemOpen
  \bibfield  {author} {\bibinfo {author} {\bibfnamefont {R.~V.}\ \bibnamefont
  {Sole}},\ }\href@noop {} {\emph {\bibinfo {title} {Phase transitions}}}\
  (\bibinfo  {publisher} {Princeton University Press},\ \bibinfo {year}
  {2011})\BibitemShut {NoStop}%
\bibitem [{\citenamefont {Mirlin}\ and\ \citenamefont
  {Fyodorov}(1991)}]{Mirlin1991}%
  \BibitemOpen
  \bibfield  {author} {\bibinfo {author} {\bibfnamefont {A.~D.}\ \bibnamefont
  {Mirlin}}\ and\ \bibinfo {author} {\bibfnamefont {Y.~V.}\ \bibnamefont
  {Fyodorov}},\ }\href {\doibase
  http://dx.doi.org/10.1016/0550-3213(91)90028-V} {\bibfield  {journal}
  {\bibinfo  {journal} {Nuclear Physics B}\ }\textbf {\bibinfo {volume}
  {366}},\ \bibinfo {pages} {507 } (\bibinfo {year} {1991})}\BibitemShut
  {NoStop}%
\bibitem [{\citenamefont {Monthus}\ and\ \citenamefont
  {Garel}(2009{\natexlab{b}})}]{GarelCayleyTree2009}%
  \BibitemOpen
  \bibfield  {author} {\bibinfo {author} {\bibfnamefont {C.}~\bibnamefont
  {Monthus}}\ and\ \bibinfo {author} {\bibfnamefont {T.}~\bibnamefont
  {Garel}},\ }\href {http://stacks.iop.org/1751-8121/42/i=7/a=075002}
  {\bibfield  {journal} {\bibinfo  {journal} {Journal of Physics A:
  Mathematical and Theoretical}\ }\textbf {\bibinfo {volume} {42}},\ \bibinfo
  {pages} {075002} (\bibinfo {year} {2009}{\natexlab{b}})}\BibitemShut
  {NoStop}%
\bibitem [{\citenamefont {Slevin}\ and\ \citenamefont
  {Ohtsuki}(1999)}]{slevin-ohtsuki-prl99}%
  \BibitemOpen
  \bibfield  {author} {\bibinfo {author} {\bibfnamefont {K.}~\bibnamefont
  {Slevin}}\ and\ \bibinfo {author} {\bibfnamefont {T.}~\bibnamefont
  {Ohtsuki}},\ }\href {\doibase 10.1103/PhysRevLett.82.382} {\bibfield
  {journal} {\bibinfo  {journal} {Phys. Rev. Lett.}\ }\textbf {\bibinfo
  {volume} {82}},\ \bibinfo {pages} {382} (\bibinfo {year} {1999})}\BibitemShut
  {NoStop}%
\bibitem [{\citenamefont {Rodriguez}\ \emph {et~al.}(2011)\citenamefont
  {Rodriguez}, \citenamefont {Vasquez}, \citenamefont {Slevin},\ and\
  \citenamefont {R\"omer}}]{rodriguez-etal-prb11}%
  \BibitemOpen
  \bibfield  {author} {\bibinfo {author} {\bibfnamefont {A.}~\bibnamefont
  {Rodriguez}}, \bibinfo {author} {\bibfnamefont {L.}~\bibnamefont {Vasquez}},
  \bibinfo {author} {\bibfnamefont {K.}~\bibnamefont {Slevin}}, \ and\ \bibinfo
  {author} {\bibfnamefont {R.}~\bibnamefont {R\"omer}},\ }\href {\doibase
  10.1103/PhysRevB.84.134209} {\bibfield  {journal} {\bibinfo  {journal} {Phys.
  Rev. B}\ }\textbf {\bibinfo {volume} {84}},\ \bibinfo {pages} {134209}
  (\bibinfo {year} {2011})}\BibitemShut {NoStop}%
\bibitem [{\citenamefont {Ueoka}\ and\ \citenamefont
  {Slevin}(2014)}]{Slevin2014}%
  \BibitemOpen
  \bibfield  {author} {\bibinfo {author} {\bibfnamefont {Y.}~\bibnamefont
  {Ueoka}}\ and\ \bibinfo {author} {\bibfnamefont {K.}~\bibnamefont {Slevin}},\
  }\href {\doibase 10.7566/JPSJ.83.084711} {\bibfield  {journal} {\bibinfo
  {journal} {Journal of the Physical Society of Japan}\ }\textbf {\bibinfo
  {volume} {83}},\ \bibinfo {pages} {084711} (\bibinfo {year} {2014})},\
  \Eprint {http://arxiv.org/abs/http://dx.doi.org/10.7566/JPSJ.83.084711}
  {http://dx.doi.org/10.7566/JPSJ.83.084711} \BibitemShut {NoStop}%
\bibitem [{\citenamefont {Hsu}\ and\ \citenamefont
  {Valles}(1995)}]{HsuVallesPRL95}%
  \BibitemOpen
  \bibfield  {author} {\bibinfo {author} {\bibfnamefont {S.-Y.}\ \bibnamefont
  {Hsu}}\ and\ \bibinfo {author} {\bibfnamefont {J.~M.}\ \bibnamefont {Valles},
  \bibfnamefont {Jr.}},\ }\href@noop {} {\bibfield  {journal} {\bibinfo
  {journal} {Phys. Rev. Lett.}\ }\textbf {\bibinfo {volume} {74}},\ \bibinfo
  {pages} {2331} (\bibinfo {year} {1995})}\BibitemShut {NoStop}%
\bibitem [{\citenamefont {Kravchenko}\ \emph {et~al.}(1996)\citenamefont
  {Kravchenko}, \citenamefont {Simonian}, \citenamefont {Sarachik},
  \citenamefont {Mason},\ and\ \citenamefont {Furneaux}}]{kravchenkoPRL96}%
  \BibitemOpen
  \bibfield  {author} {\bibinfo {author} {\bibfnamefont {S.~V.}\ \bibnamefont
  {Kravchenko}}, \bibinfo {author} {\bibfnamefont {D.}~\bibnamefont
  {Simonian}}, \bibinfo {author} {\bibfnamefont {M.~P.}\ \bibnamefont
  {Sarachik}}, \bibinfo {author} {\bibfnamefont {W.}~\bibnamefont {Mason}}, \
  and\ \bibinfo {author} {\bibfnamefont {J.~E.}\ \bibnamefont {Furneaux}},\
  }\href@noop {} {\bibfield  {journal} {\bibinfo  {journal} {Phys. Rev. Lett.}\
  }\textbf {\bibinfo {volume} {77}},\ \bibinfo {pages} {4938} (\bibinfo {year}
  {1996})}\BibitemShut {NoStop}%
\end{thebibliography}%


\begin{thebibliography}{1}%
\makeatletter
\providecommand \@ifxundefined [1]{%
 \@ifx{#1\undefined}
}%
\providecommand \@ifnum [1]{%
 \ifnum #1\expandafter \@firstoftwo
 \else \expandafter \@secondoftwo
 \fi
}%
\providecommand \@ifx [1]{%
 \ifx #1\expandafter \@firstoftwo
 \else \expandafter \@secondoftwo
 \fi
}%
\providecommand \natexlab [1]{#1}%
\providecommand \enquote  [1]{``#1''}%
\providecommand \bibnamefont  [1]{#1}%
\providecommand \bibfnamefont [1]{#1}%
\providecommand \citenamefont [1]{#1}%
\providecommand \href@noop [0]{\@secondoftwo}%
\providecommand \href [0]{\begingroup \@sanitize@url \@href}%
\providecommand \@href[1]{\@@startlink{#1}\@@href}%
\providecommand \@@href[1]{\endgroup#1\@@endlink}%
\providecommand \@sanitize@url [0]{\catcode `\\12\catcode `\$12\catcode
  `\&12\catcode `\#12\catcode `\^12\catcode `\_12\catcode `\%12\relax}%
\providecommand \@@startlink[1]{}%
\providecommand \@@endlink[0]{}%
\providecommand \url  [0]{\begingroup\@sanitize@url \@url }%
\providecommand \@url [1]{\endgroup\@href {#1}{\urlprefix }}%
\providecommand \urlprefix  [0]{URL }%
\providecommand \Eprint [0]{\href }%
\providecommand \doibase [0]{http://dx.doi.org/}%
\providecommand \selectlanguage [0]{\@gobble}%
\providecommand \bibinfo  [0]{\@secondoftwo}%
\providecommand \bibfield  [0]{\@secondoftwo}%
\providecommand \translation [1]{[#1]}%
\providecommand \BibitemOpen [0]{}%
\providecommand \bibitemStop [0]{}%
\providecommand \bibitemNoStop [0]{.\EOS\space}%
\providecommand \EOS [0]{\spacefactor3000\relax}%
\providecommand \BibitemShut  [1]{\csname bibitem#1\endcsname}%
\let\auto@bib@innerbib\@empty
\bibitem [{\citenamefont {Weiss}(2013)}]{dataCppWeissBook}%
  \BibitemOpen
  \bibfield  {author} {\bibinfo {author} {\bibfnamefont {M.~A.}\ \bibnamefont
  {Weiss}},\ }\href@noop {} {\emph {\bibinfo {title} {Data Structures $\&$
  Algorithm Analysis in C++}}}\ (\bibinfo  {publisher} {Prentice Hall; 4
  edition},\ \bibinfo {year} {2013})\BibitemShut {NoStop}%
\end{thebibliography}%

\end{document}


\title{Supplemental notes for: ``Strong-disorder renormalization group study
of the Anderson localization transition in three and higher dimensions''}

\author{H. Javan Mard}

\affiliation{Department of Physics and National High Magnetic Field Laboratory,
Florida State University, Tallahassee, FL 32306}

\author{José A. Hoyos}

\affiliation{Instituto de Física de São Carlos, Universidade de São Paulo, C.P.
369, São Carlos, SP 13560-970, Brazil}

\author{E. Miranda}

\affiliation{Instituto de Física Gleb Wataghin, Unicamp, R. Sérgio Buarque de
Holanda, 777, Campinas, SP 13083-859, Brazil}

\author{V. Dobrosavljevi\'{c}}

\affiliation{Department of Physics and National High Magnetic Field Laboratory,
Florida State University, Tallahassee, FL 32306}

\date{\today}

\maketitle

\section{SDRG approach computes the conductance exactly}

It is easy to prove that the SDRG transformations derived in our previous
work leads to compute the exact two point conductance in the Andersom
model. If we integrate out a site, say $m$, during SDRG process while
keeping the Green's function conserved between two arbitrary sites
$i,$ and $j$:

\begin{equation}
G(i,j)=\frac{1}{Z}\int D[\phi_{k}(\tau),\phi_{k}^{*}(\tau)]\ \phi_{i}(\tau)\ \phi_{j}(\tau)\ e^{-S[\phi_{k}(\tau),\phi_{k}^{*}(\tau)]},\label{eq:Green'sfxn}
\end{equation}

where 
\[
D[\phi_{k}(\tau),\phi_{k}^{*}(\tau)]=\prod_{k=1}^{n}d\phi_{k}(\tau)\ d\phi_{k}^{*}(\tau),
\]

and 
\begin{eqnarray}
S[\phi_{k}(\tau),\phi_{k}^{*}(\tau)] & = & -\int_{0}^{\beta}d\tau[\sum_{k}\phi_{k}^{*}(\tau)\{\frac{\partial}{\partial\tau}-\mu+\varepsilon_{k}\}\phi_{k}(\tau)\nonumber \\
 &  & +\sum_{k,l}\phi_{k}^{*}t_{kl}\phi_{l}].\label{eq:Action}
\end{eqnarray}

Using the frequency representation, 

\begin{eqnarray}
S[\phi_{k}(\omega),\phi_{k}^{*}(\omega)] & = & -[\sum_{k,n}\phi_{k}^{*}(\omega_{n})\{i\omega_{n}-\mu+\varepsilon_{k}\}\phi_{k}(\omega_{n})\nonumber \\
 &  & +\sum_{k,l}\phi_{k}^{*}(\omega_{n})\ t_{kl}\phi_{l}(\omega_{n})].\label{eq:actionOmega}
\end{eqnarray}

At $\omega=\mu=0$ , we can simplify the above equation to $S=-[\sum_{k}\phi_{k}^{*}\{\varepsilon_{k}\}\phi_{k}+\sum_{k,l}(\phi_{k}^{*}t_{kl}\phi_{l}+h.c.)]$.
Then, those terms in S associated to $m$ can be rewrite as:

\[
-\phi_{m}^{*}\varepsilon_{m}\phi_{m}-\sum_{k}(\phi_{m}^{*}t_{mk}\phi_{k}+h.c.)=
\]

\begin{equation}
\,-\varepsilon_{m}[(\phi_{m}^{*}+\sum_{k}\frac{t_{mk}}{\varepsilon_{m}}\phi_{k}^{*})(\phi_{m}+\sum_{k}\frac{t_{mk}}{\varepsilon_{m}}\phi_{k})]+\sum_{kl}\phi_{k}^{*}\frac{t_{mk}t_{ml}}{\varepsilon_{m}}\phi_{l}\label{eq:Mterms}
\end{equation}

The first term in Eq. 4 will be canceled out by doing the same calculation
in denominator of $G(i,j)$( we assume $\varepsilon_{m}$ is nonzero.),
but the second term leads to renormalize the hopping term between
sites $k$ and $l$:

\begin{equation}
\widetilde{t}_{kl}=t_{kl}-\frac{t_{mk}t_{ml}}{\varepsilon_{m}}\label{eq:t-renormalize}
\end{equation}

This is the same decimation rules for $\varepsilon$ , and $t$-decimations
of our SDRG approach.

\textbf{How about t-decimations?}

Decimation of a bond also can be described as integrating out its
two neighbor sites one by one in the purpose of conservation of two
point Green's function. For example, assume we first integrate out
the site 2 and then site 3. After first decimation,we have

\[
\widetilde{t}_{kl}=t_{kl}-\frac{t_{2k}t_{2l}}{\varepsilon_{2}}
\]

then by decimating the site 3 , 

\begin{eqnarray*}
\widetilde{\widetilde{t}}_{kl} & = & \widetilde{t}_{kl}-\frac{\widetilde{t}_{3k}\widetilde{t}_{3l}}{\widetilde{\varepsilon}_{3}}=t_{kl}-\frac{t_{2k}t_{2l}}{\varepsilon_{2}}-\frac{(t_{3k}-\frac{t_{23}t_{2k}}{\varepsilon_{2}})(t_{3l}-\frac{t_{23}t_{2l}}{\varepsilon_{2}})}{\varepsilon_{3}-\frac{t_{23}^{2}}{\varepsilon_{2}}}\\
\\
\\
 &  & =t_{kl}+\frac{\varepsilon_{3}t_{2k}t_{2l}+t_{23}(t_{2k}t_{3l}+t_{2l}t_{3k})+\varepsilon_{2}t_{3k}t_{3l}}{t_{23}^{2}-\varepsilon_{2}\varepsilon_{3}}.
\end{eqnarray*}

This is exactly what we use as SDRG decimation rules for any dimensions.

\section{\textcolor{black}{The modified SDRG algorithm }}

1- $N$ energy sites are randomly chosen from a uniform distribution.
In the case of ER random graph, two sites are conencted with a probability
less than $p=<k>/(n-1)$, thus we can generate a graph either by checking
this condition for each pair of sites or by randomly choosing $m=pn$
of pairs of sites. The latter method is useful specially for very
large n. Since we only keep $k_{max}$ bonds per site, the required
memory space is $k_{max}n$ and it takes $\mathcal{O}(mn)$ to create
the Hamiltonian matrix. We recall that $m\ll n$ for a sparse graph. 

2- In the case of ER random graph, we use Dijkstra algorithm to find
the shortest path between sites, but only those sites that they have
a relative distance $l_{ER}=\ln(N)/\ln(<k>)$ will be chosen to be
attached to the leads. The worst running time of this algorithm is
known $\mathcal{O}(m\log n)$ to determine all shortest paths from
one reference site. Moreover, above the percolation threshed , we
are only interested in the giant component of graph and one can ignore
sites excluded form this component (doing that decrease the number
of considered sites by almost \%5 in ER graph with $<k>=3$ which
is large number when n is very large.).

3-Decimation process: We search for the maximum of $\{\varepsilon_{i},t(i,j)\}$
, and decimate them as it is explained in the main text. Performing
a SDRG decimation causes the adjacency list shrinks at least by one
site in each step , and gives a faster running time for the rest of
code in return. It is important that the sites attached to the leads,
and their bonds will not be decimated during SDRG decimation, but
their value still can be updated if a decimation happens in their
neighbor.

Decimation of n sites consists of two parts: 

$\quad$3-1 Search for the maximum of energy terms: This will take
$\mathcal{O}(\log n)$ if one use the Heap data structure to sort
$\varepsilon_{i}$ and $t_{ij}$ values separately(More details on
heap data structure, and its implementation in C++ programming language
can be found in many books such as an interesting book written by
Mark A. Weiss \cite{dataCppWeissBook}) .

$\quad$3-2 Renormalization the neighbors: the number of neighbors
are limited to $k_{max}$ so there are $k_{max}$ onsite energies
and $k_{max}^{2}$ hopping terms :$\mathcal{O}(k_{max}^{2}\log n)$

4- After decimating a site or bond if the coordination number of a
particular site say $j$ , exceeds $k_{max}$ , we replace the weakest
bond of site j with the new bond, otherwise we ignore it.

Therefore, the total running time of steps1-4 would be $\mathcal{O}(n\log n)$
. 

5- Steps 1-4 can be repeated for different realization of initial
disorder until achieving the desired accuracy.

Fig. 3 presents the efficiency of SDRG modified algorithm used for
a 3D lattice in a range of system sizes accessible for all methods.

\section{Computation of critical exponent $\nu$ in $d>3$}

As we discussed earlier, the modified SDRG algorithm only depends
on the number of sites in the lattice. Therefore, it is possible to
take advantage of this method and compute $\nu$ in higher dimensions.
However, the number of sites in higher dimensions grows faster with
the lattice size and one should be concerned about the accessible
computer memory and time, and the effect of $k_{max}$ when it gets
closer to the initial coordination number. 

\begin{figure}[H]
\includegraphics[width=1\columnwidth]{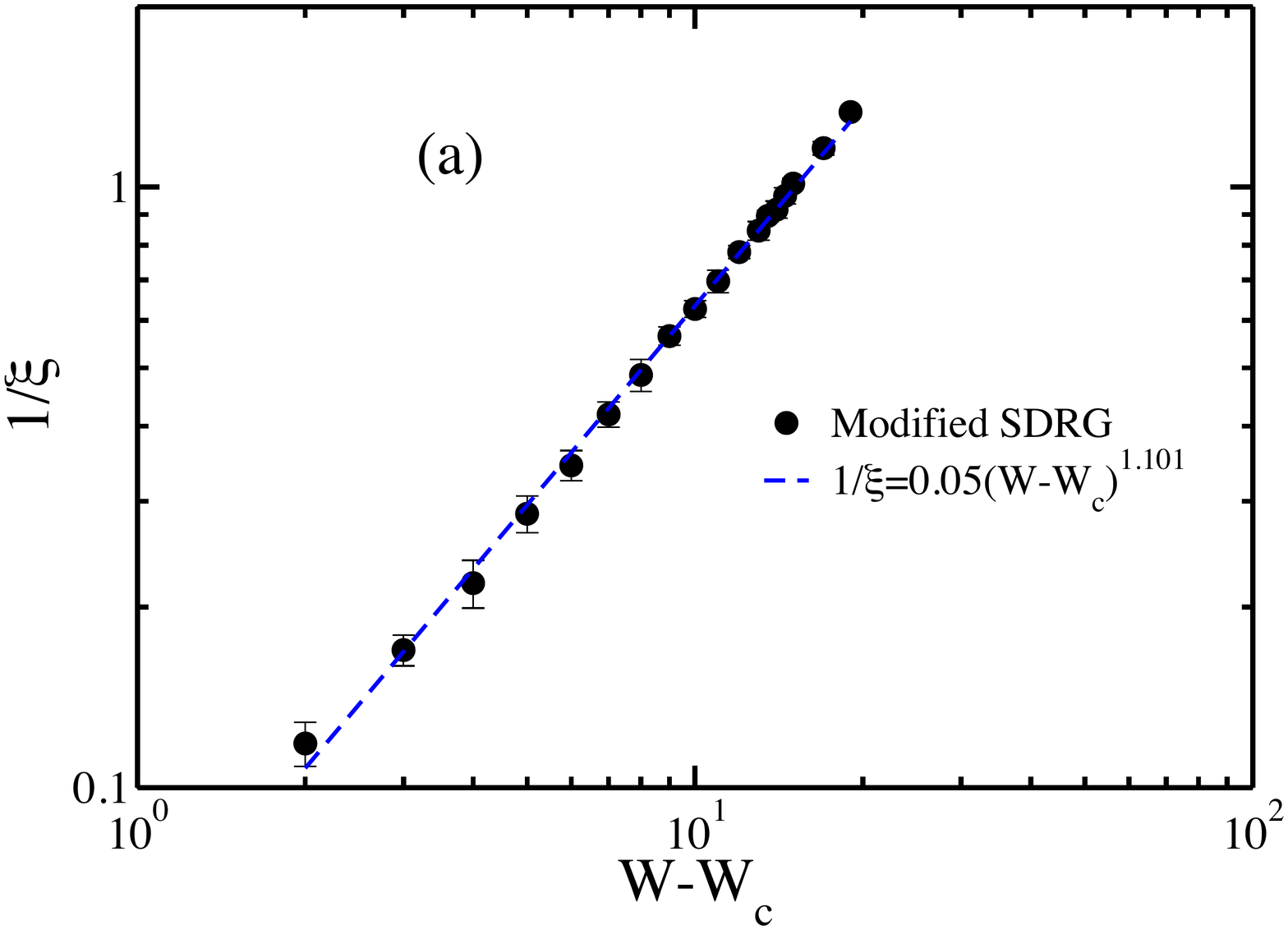}

\includegraphics[width=1\columnwidth]{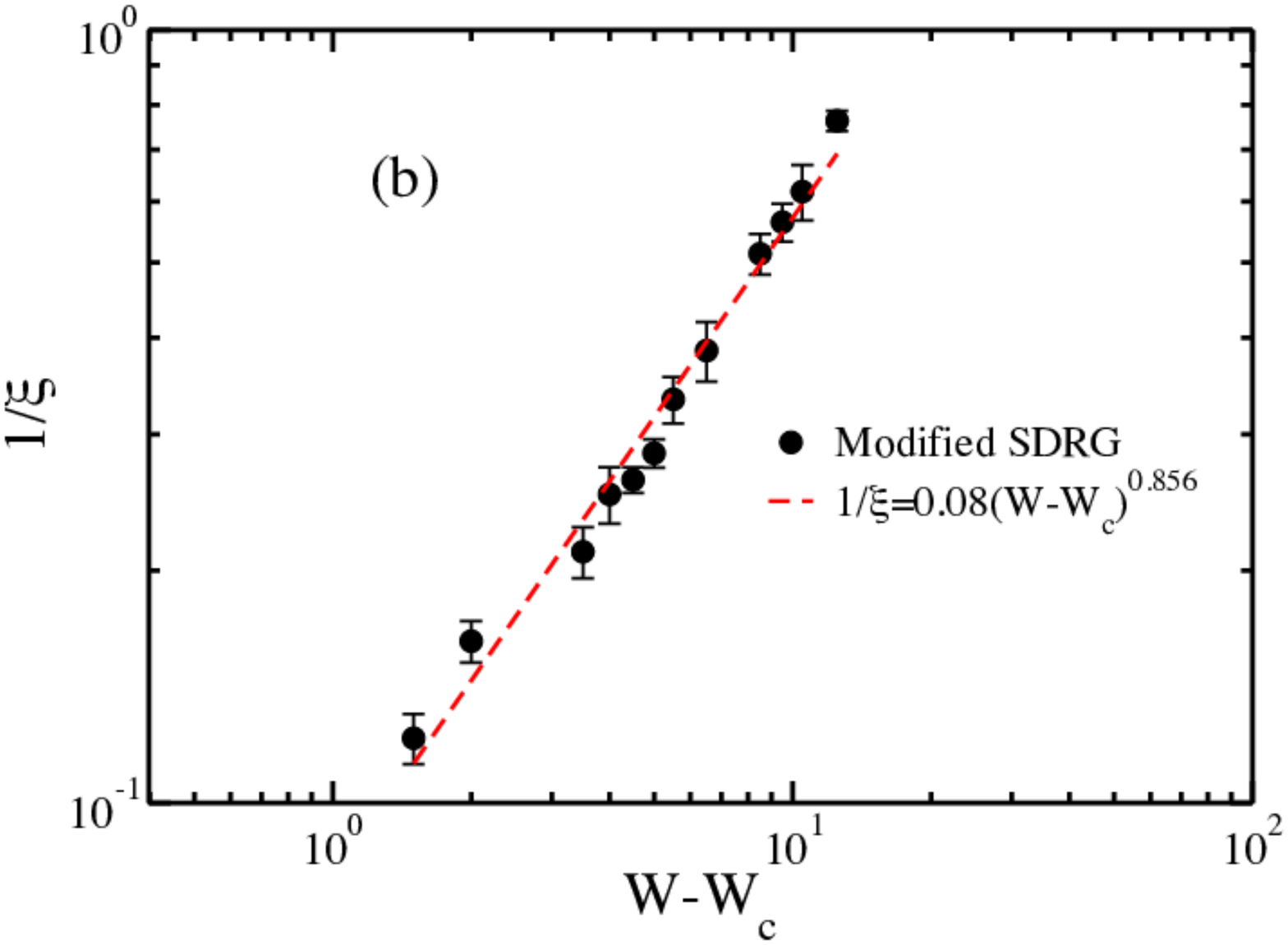}

\includegraphics[width=1\columnwidth]{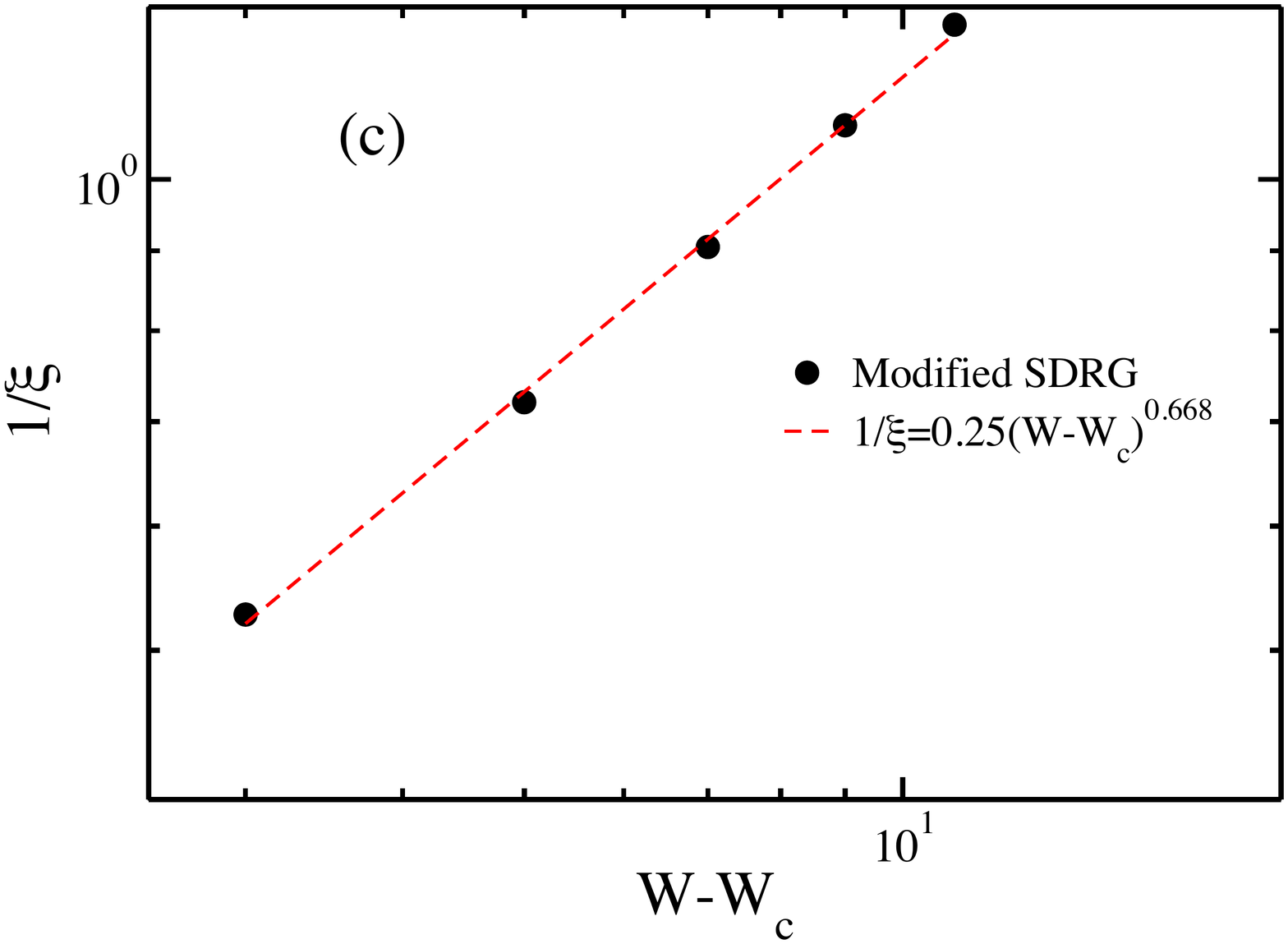}

\protect\caption{The critical exponent of the localization length is computed for a
hyper-cubic lattice with $d=4,6,$ and $10$. (a) The critical exponent
$\nu$ estimated as $\nu_{4D}=1.10\pm0.09$ and $W_{c}=22\pm0.5$
. Here, we studied the disorder strength from 5 to 41, and system
sizes $L=6,8,10,$and $12$. (b) In six dimension, the critical exponent
computed as $\nu_{6D}=0.86$ and $W_{c}=51.5\pm0.5$ . Here, $W$
was changed from 20 to 70 and $L$ from $6$ to $12$ . (c) In 10D
lattice, we used system sizes $L=3,4,5$ and disorder strength from
$80t$ to $110t$ . Our studies shows $W_{c}=97\pm0.5$ and $\nu_{10D}=0.68\pm0.06$
.}
\end{figure}

\section{Mirror Symmetry Analysis:}

The Mirror Symmetry Range (MSR) is defined as the range of $g/g_{c}$
where mirror symmetry holds. Although this range can be simply estimated
approximately just by looking or inverting the scaling function in
one phase onto another one, but we measure MSR in a more well-defined
approach as follows: 

1-We plot $t=\ln(g/g_{c})$ versus $y=\pm x^{1/\nu}$ where $x=L/\xi$
where the plus sign corresponds to the localized phase and the minus
to the delocalized phase. Let's call the resulting function $t(y).$

2- In the presence of symmetry, $t(y)$ is an odd function. Thus,
we separate the odd and even parts of function t(y) and find where
the even part value becomes significant. In other words,

\begin{equation}
t_{odd}(y)=\frac{1}{2}[t(y)-t(-y)],\ t_{even}(y)=\frac{1}{2}[t(y)+t(-y)]\label{eq:evenOdd}
\end{equation}

\[
t_{even}(y^{*})=\alpha t_{odd}(y^{*}),
\]

where $\alpha$ determines the error of our estimation. For our convenience,
we fit $t(y)$ to a polynomial function in each phase separately before
determining its odd and even parts. 

\begin{figure}[H]
\includegraphics[width=0.49\columnwidth]{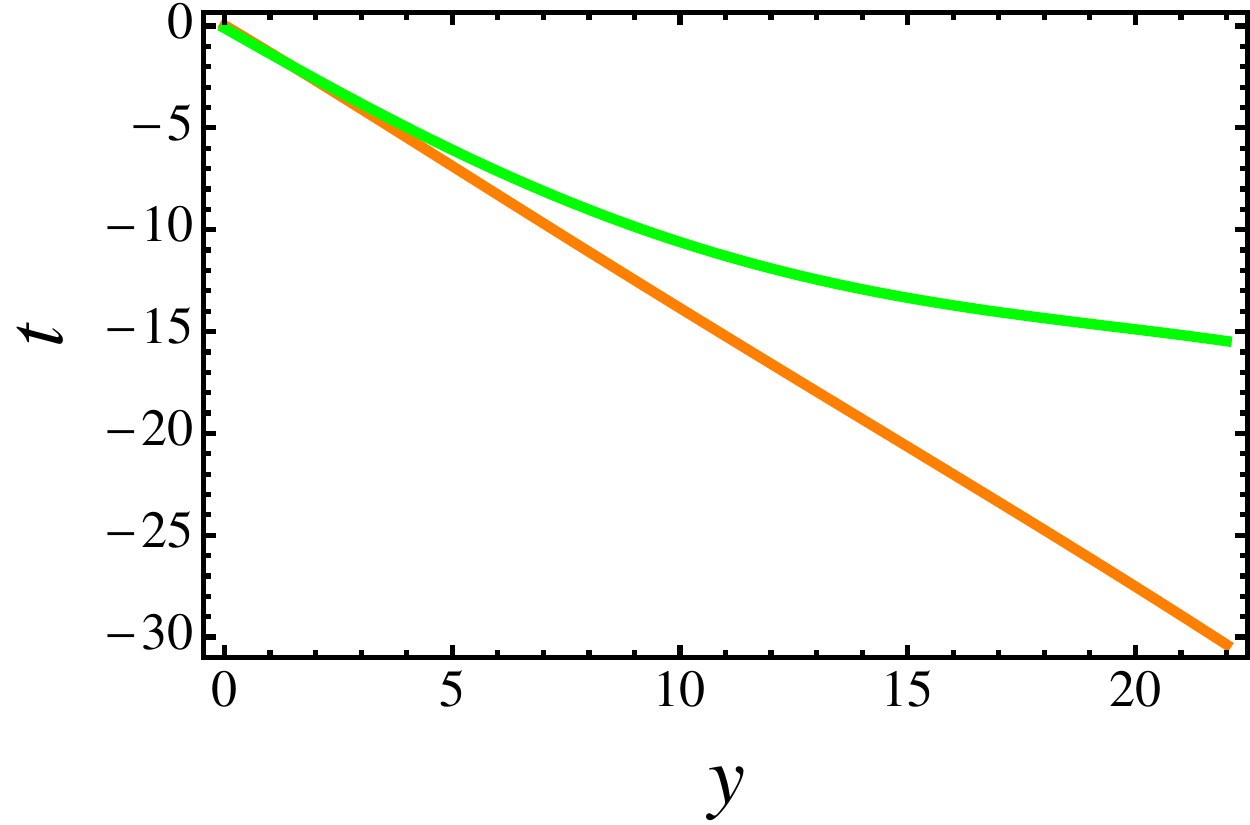} \includegraphics[width=0.45\columnwidth]{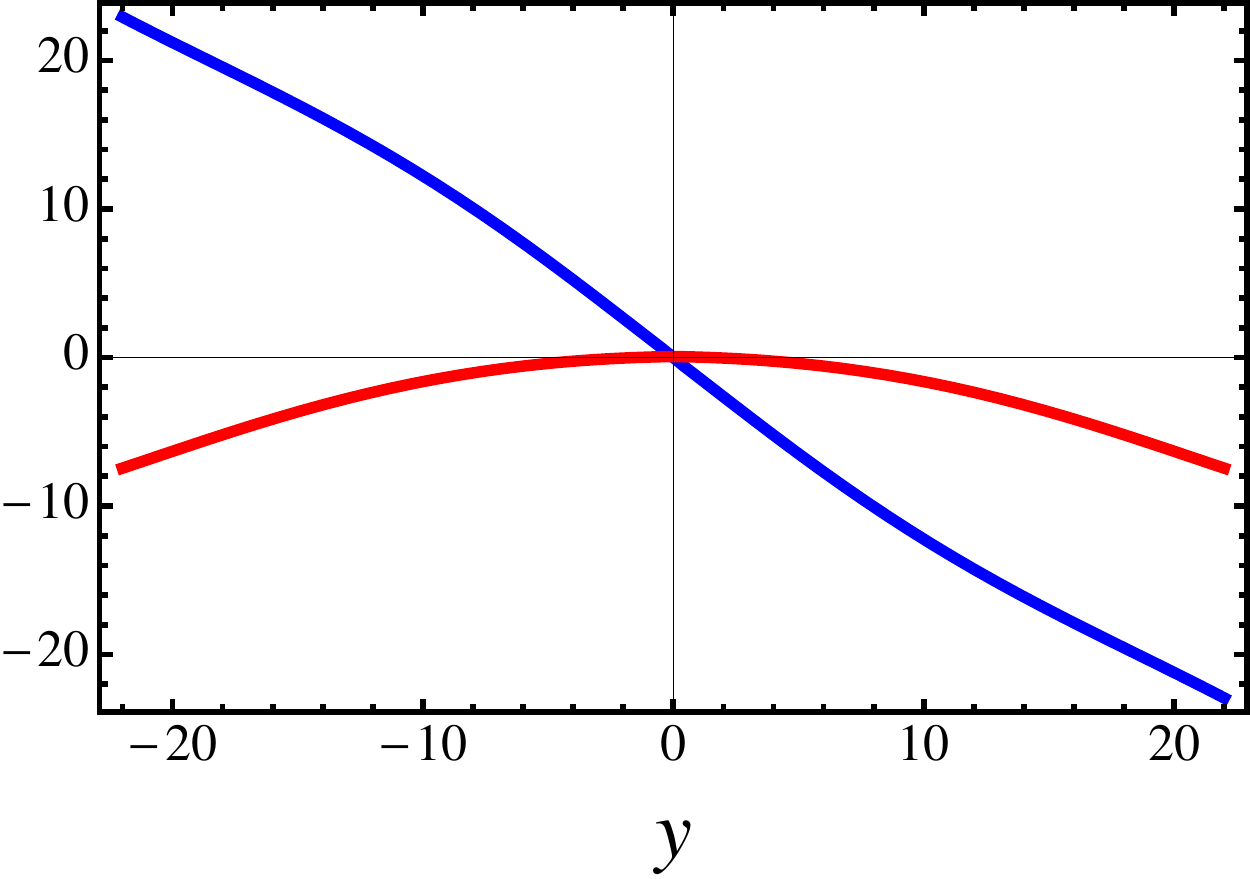}

\protect\caption{The final steps to measure the mirror symmetry range is presented
for 4D lattice equal 4.34 (Left) $t(y)$ is plotted from fitting the
modified SDRG results onto a 5th order polynomial function in each
phase separately. Here the orange color corresponds to the localized
and the green color to the delcoalized phase (inverted to the positive
$y$). (Right) The even and odd parts of function $t(y)$ is plotted
seperately.}
\end{figure}

In contrast to the $2+\epsilon$ expansion theory which only predicts
very small symmetry range, we find a significant range of mirror symmetry
in 3D and higher. To see $2+\epsilon$ expansion theory prediction
for MSR, one can compute the scaling function $g$ close to $d=2$,
i.e. for $\epsilon=d-2\ll1,$ we have the perturbative result:
\begin{equation}
\beta(g)\approx\epsilon-\frac{a}{g}+\cdots,\label{eq:beta2epsilon}
\end{equation}
or
\begin{equation}
\frac{1}{\epsilon}\int_{g(\ell)}^{g(L)}\frac{dg}{g-g_{c}}=\ln(L/\ell),\label{eq:integralBeta}
\end{equation}
where $g_{c}=a/\epsilon$, $\ell$ is the mean-free path, and $L$
is the system size; we assumed that we are on the metallic side, so
$g(\ell)>g_{c}$. Let us denote $g(\ell)=g_{o}$, and $\delta g_{o}=g_{o}-g_{c}.$
We obtain:

\[
g(L)=g_{c}+\delta g_{o}(L/\ell)^{\epsilon},
\]
or 
\[
g(L)=F(L/\xi),
\]
where $\xi=\ell\delta g^{-\nu}$, with $\nu=\epsilon^{-1}.$, and
the ``metallic'' branch of the scaling function of the form 
\[
F_{met.}(x)=g_{c}+x^{\epsilon}.
\]

For large system sizes $g(L)\gg g_{c}$, and from the definition of
the conductivity $g(L)=L^{\epsilon}\sigma$, we obtain
\[
\sigma=\delta g_{o}^{\mu},
\]
with $\mu=\epsilon\nu$. Note also that the quantity $\delta g_{o}$
measures the distance to the transition, i.e. $\delta g_{o}\sim(W-W_{c})$,
etc. Note also that for $\epsilon\ll1$, this calculation is sufficient
on the entire\textbf{ }metallic side, since higher order terms in
$g^{-1}$ are negligible. 

The same argument does not hold on the insulating side, since there
$g(L)$ flows to zero, hence the higher order terms become large (in
magnitude, as $\beta<0$ there). Still, sufficiently close to the
transition, the above ``perturbative'' form of the beta function
remains sufficient, and we can repeat the same integration procedure
(except now $g(L)<g_{c}$). We find
\[
g(L)=g_{c}-(L/\xi)^{\epsilon},
\]
with $\xi=\ell|\delta g_{o}|^{-\nu}.$ The corresponding ``insulating''
branch of the scaling function is 
\[
F_{ins.}(x)=g_{c}-x^{\epsilon}.
\]

If we define the quantity $y=\pm x^{\epsilon}$, as usual, we can
express both branches as parts of the same scaling function
\[
F(y)=g_{c}+y.
\]
This is an exact result for the scaling function in the regime where
the perturbative form is valid. 

Now we define the function 
\begin{equation}
t(y)=\ln(g/g_{c})=\ln(1+\epsilon y).\label{eq:tMSR}
\end{equation}
Obviously, this function has mirror symmetry, only for $\epsilon y\ll1.$
In terms of the MSR, this means only a very small range. We can derive
a more formal range using separating even and odd parts of $t(y)$
. If we use $\alpha<<1$ as accuracy of our results in Eq. \ref{eq:evenOdd}

\begin{eqnarray*}
\frac{t_{even}(y^{*})}{t_{odd}(y^{*})} & = & \alpha\longrightarrow\ln(1+\epsilon y^{*})(1-\epsilon y^{*})=\\
 &  & \alpha\ln(\frac{1+\epsilon y^{*}}{1-\epsilon y^{*}})
\end{eqnarray*}

\begin{equation}
(1+\epsilon y^{*})^{\alpha+1}=(1-\epsilon y^{*})^{\alpha-1}\label{eq:msrEquation}
\end{equation}

Solving the last equation gives two solutions $y^{*}=0$ and $y^{*}=2\alpha/\epsilon$.
By putting back $y^{*}$ in Eq. \ref{eq:tMSR} , and solving for $g/g_{c}$
we find $MSR=1+2\alpha$ which is a number of order of one.

\section{the role of $k_{max}$ in the accuracy of results}

We introduced $k_{max}$ as an upper cut-off for the number of bonds
we store during SDRG decimation. Although this algorithm speeds up
the running time(see Fig. 3) , this idea is a heuristic argument and
one expects an approximated results as we throw away more bonds no
matter how weak are. One can look at this parameter as a control tool
to adjust the accuracy of results as desired. In the extreme limits,
when $k_{max}$ approaches to $N$ we get exact results and we loose
significant information if we push it to very small numbers. In figure
4, we present the sensitivity of our results to the value of $k_{max}$
for a cubic lattice. This figure does not necessarily imply anything
about the best choice of $k_{max}$ for other dimensions and even
for larger system sizes. One safer approach might be to vary $k_{max}$
as $N$ increases and dimensionality changes. 

\begin{figure}[h]
\includegraphics[angle=-90,width=1\columnwidth]{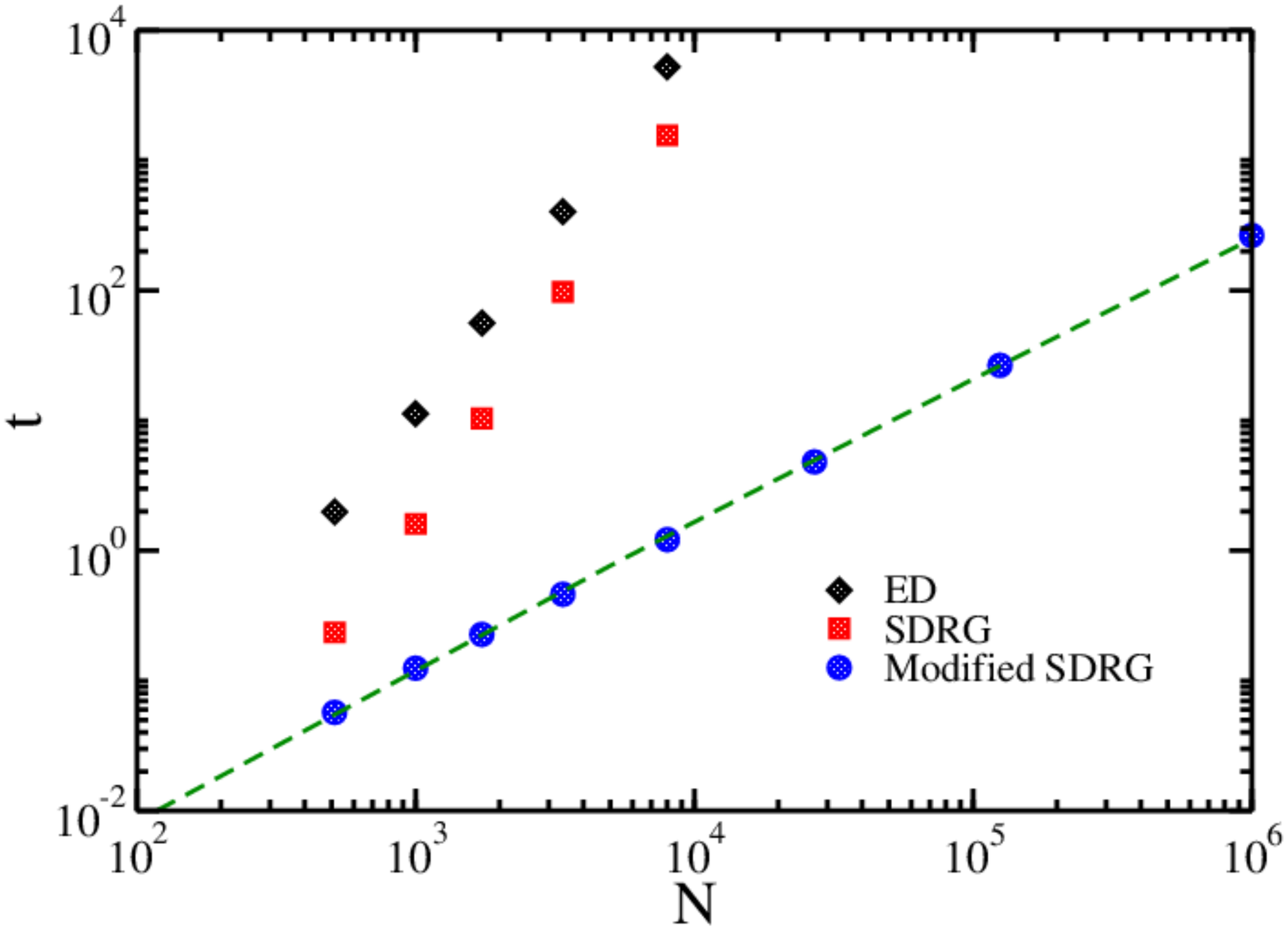}

\protect\caption{Computational time of the algorithm, $t$ (in seconds in a 3.4GHz
processor) for single realization of disorder as a function of the
size of the cubic lattice, $N=L^{3}$, in a log-log scale for three
different methods. The dashed line describes our theoretical prediction,
$t\sim N\log N$. Here, the disorder value of the sample is $W=17$.}
\end{figure}

\begin{figure}[H]
\includegraphics[width=1\columnwidth]{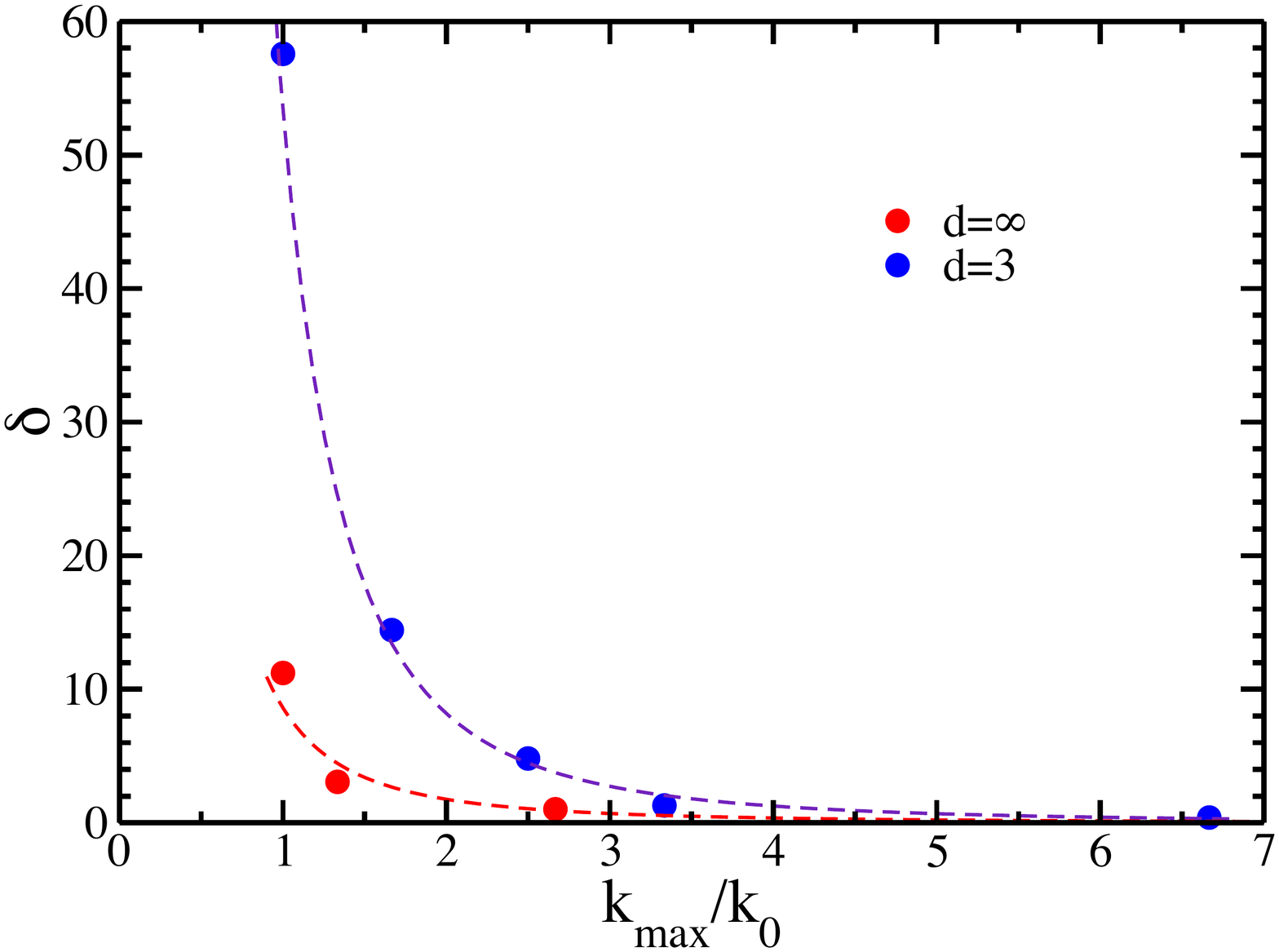}

\protect\caption{The produced error $\delta=\frac{|\nu-\nu_{exact}|}{\nu_{exact}}\times100$
in the computed critical exponent $\nu$ by varying $k_{max}$ for
3D lattice and ER random graph. The same range of disorder strength
from $8$ to $24$ is used for both models. The maximum system sizes
in 3D lattice is $L=20$ while For infinite dimensional limit, we
used the ER random graph samples with $<k>=3.0$, the same range of
disorder as cubic lattice, and site number $N=3^{l_{ER}}$ where $l_{ER}=4,5,6,7$.
Here, the $k_{max}$ is normalized by the maximum initial coordination
number, $k_{0}$. }
\end{figure}

\bibliographystyle{apsrev4-1}
\bibliography{bibliography}